\documentclass[12pt]{article}
\pdfoutput=1

\usepackage{graphicx}
\usepackage{amssymb}
\usepackage{amsmath}
\usepackage{esint}

\usepackage{bm}

\usepackage{bm}

\newcommand{\rf}[1]{(\ref{#1})}
\newcommand{\beq}{\begin{equation}}
\newcommand{\beql}[1]{\beq\label{#1}}
\newcommand{\eeq}{\end{equation}}
\newcommand{\bea}{\begin{eqnarray}}
\newcommand{\eea}{\end{eqnarray}}

\newcommand{\e}{\mbox{e}}

\newcommand{\lam}{\lambda}

\newcommand{\ep}{\varepsilon}

\newcommand{\Om}{\Omega}
\newcommand{\del}{\delta}
\newcommand{\Del}{\Delta}

\newcommand{\kp}{\kappa}

\newcommand{\tr}{\mathrm{tr}\,}
\newcommand{\ra}{\rangle}
\newcommand{\la}{\langle}
\newcommand{\braket}[1]{\la {#1} \ra}
\newcommand{\ket}[1]{| {#1} \ra}
\newcommand{\bra}[1]{\la {#1} |}

\newcommand{\mi}{\!-\!}
\newcommand{\equ}{\!=\!}
\newcommand{\plu}{\!+\!}

\newcommand{\cM}{{\cal M}}

\newcommand{\cT}{{\cal T}}

\newcommand{\cN}{{\cal N}}

\newcommand{\cZ}{{\cal Z}}

\newcommand{\tS}{{\tilde{S}}}

\newcommand{\tM}{{\tilde{M}}}
\newcommand{\tP}{{\tilde{P}}}
\newcommand{\trho}{{\tilde{\rho}}}

\begin{document}

\begin{center}
\vspace{24pt}
{ \large \bf The transfer matrix in four-dimensional CDT}

\vspace{30pt}

{\sl J. Ambj\o rn}$\,^{a}$,
{\sl J. Gizbert-Studnicki}$\,^{b}$
{\sl A. T. G\"{o}rlich}$\,^{a,b}$
{\sl J. Jurkiewicz}$\,^{b}$,

\vspace{24pt}
{\footnotesize

$^a$~The Niels Bohr Institute, Copenhagen University\\
Blegdamsvej 17, DK-2100 Copenhagen \O , Denmark.\\

\vspace{5pt}

{ email: ambjorn@nbi.dk,~goerlich@nbi.dk}\\

\vspace{15pt}

$^b$~Institute of Physics, Jagellonian University,\\
Reymonta 4, PL 30-059 Krakow, Poland.\\

\vspace{5pt}

{ email: atg@th.if.uj.edu.pl, jakub.gizbert-studnicki@uj.edu.pl\\
~jurkiewicz@th.if.uj.edu.pl }\\

\vspace{10pt}

}
\vspace{48pt}

\end{center}


\begin{center}
{\bf Abstract}
\end{center}

\noindent
The Causal Dynamical Triangulation model of quantum gravity  (CDT)
has a transfer matrix, relating spatial geometries at adjacent 
(discrete lattice) times. The transfer matrix uniquely determines
the theory. We show that the measurements of the scale factor
of the (CDT) universe are well described by an effective 
transfer matrix where the matrix elements are labeled only 
by the scale factor.  Using computer simulations we determine the 
effective transfer matrix elements and show how they 
relate to an effective minisuperspace action at all scales.

\newpage

\section{Introduction}

Minisuperspace models of the universe provide us with 
simple quantum mechanical models of fluctuations of the 
scale factor of the universe. In the simplest models one
assumes spatial homogeneity and isotropy. Classically this 
implies that we can write 
\beq\label{j1}
ds^2= -dt^2 + a^2(t) d\Om^2
\eeq
where $d\Om^2$ is the line element of a homogeneous and isotropic 
three-dimensional space. Assuming space is compact it is $S^3$.  
Under these assumptions the dynamical variable is the scale 
factor $a(t)$ of the universe and the quantum field theory of the 
universe is reduced to quantum mechanics of a single variable $a(t)$.

The 4D Causal Dynamical Triangulation model (CDT) of quantum gravity
is by construction a (regularized) quantum field theory model 
which is compatible with spatial homogeneity and isotropy
(for reviews see \cite{CDTreviews,physrep}).
It uses the path integral formulation and assumes there exists a foliation
in (proper) time. When the average over all geometries of this kind
is performed one indeed finds that the average geometry, i.e.\ what 
one naively would think is closest to a  ``classical geometry'', can
be described by a line element of type \rf{j1}. Even more, it turns
out that to a good approximation the scale factor $a(t)$, measured
as the average of the third root of the spatial volume at (proper) time 
$t$ is well described by the simplest minisuperspace action. Contrary to the 
standard use of minisuperspace, where one {\it postulates} the reduction
to geometries described by a metric like \rf{j1} and then 
only quantizes the single degree of freedom $a(t)$, the CDT 
discussion of $a(t)$ is exact (to the extent that CDT describes quantum
gravity). The $a(t)$ entering in the CDT discussion is obtained 
by including all geometries in the path integral. It is natural 
to ask the following question: how well does 
the simplest minisuperspace action describe the CDT data generated
by Monte Carlo simulations. The tool we will use when trying to  answer this 
question is the transfer matrix.
   
The time foliation present in CDT provides us with a transfer matrix. 
The geometries considered in the (regularized) path integral are 
piecewise linear geometries constructed in such a way that at discretized
times $t_n$ the spatial slices are triangulations of $S^3$. The transfer
matrix relates a given spatial (piecewise linear) 
geometry at time $t_n$ to a given spatial geometry 
at time $t_{n+1}$. In ordinary (Euclidean) lattice field theory
reflection positiveness of the transfer matrix ensures a unitary
time evolution. 
The CDT transfer matrix has a similar property \cite{ajl2001,physrep}
\footnote{The idea of a time foliation and the fact that 
there exists a related unitary time evolution are features
which CDT shares with Ho\v rava-Lifshitz gravity \cite{horava}. However,
no  spatial higher derivative terms are explicitly added 
to the action, like in Ho\v rava-Lifshitz gravity, and 
it is possible that  fixed points of the lattice theory 
can be identified with the non-trivial UV fixed points conjectured in the 
asymptotic safety scenario suggested by Weinberg \cite{weinberg}
and investigated in \cite{kawai,reuter,reuteretc}.}.
In this article we will analyze an ``effective''  transfer matrix of CDT 
to be defined below.

\section{CDT and the CDT transfer matrix}

The use of piecewise linear geometries was introduced in the context of  
general relativity by Regge \cite{regge} as a natural tool  to 
work with a  discretized version of the  Hilbert-Einstein action,
but without the use of coordinates. The idea was to approximate 
a given smooth geometry by a continuous, piecewise linear 
geometry. This piecewise linear geometry is uniquely defined
by a triangulation where the length of the links are given. 
The observation by Regge was that the standard  Einstein-Hilbert 
action in $D$ dimensions, 
\begin{equation}
S_{HE}[g]=\frac{1}{16 \pi G}\int{d^D x \sqrt{-g}(R-2 \Lambda)}, 
\label{SHE}
\end{equation}
for such piecewise linear geometries has a geometric interpretation
as a sum over deficit angles of the $D-2$-dimensional sub-simplices
in the triangulation. While the idea of Regge was to approach 
a given classically smooth geometry by a sequence of 
suitable piecewise linear 
triangulations, a different use of piecewise linear geometries was made 
in the formalism of Dynamical Triangulations (DT) \cite{DT}. 
Although we have no mathematical rigorous definition of the 
path integral over geometries, it is natural, in analogy with the 
path integral in ordinary quantum mechanics, to assume that the 
summation over geometries will involve not only smooth geometries
but all continuous geometries. A subclass of these geometries is
the piecewise linear geometries and a further subclass is the 
piecewise linear geometries defined by triangulations obtained 
by gluing together  equilateral $D$-simplices such that they form a
manifold of fixed topology. In DT the assumption 
is that this set of geometries is in a suitable sense dense 
in the set of continuous geometries when we take the link length
$a$ to zero, and in this way the link length $a$ will act as an 
ultraviolet cutoff, just like in ordinary lattice field theory.
Further, the natural signature of space-time in the DT formulation
is Euclidean. We will assume this is the case in the rest of 
this article. That this approach works in principle, i.e. that it is able 
to reproduce a continuum quantum field theory which is 
diffeomorphism invariant, is well documented for $D=2$. 
Two-dimensional Euclidean quantum gravity coupled to a 
conformal field theories  with $c \leq 1$ can be solved analytically 
both in the continuum \cite{KPZ, Distler:1988jt} and using the formalism of 
DT \cite{2d-DT} and agreement is found.
 
In higher dimensions  DT was studied numerically, using Monte Carlo
simulations both in three dimensions \cite{3DEuclid} and four dimensions 
\cite{4DEuclid}. However, no convincing continuum limit has been obtained 
so far in higher dimensions \cite{firstorder}, 
and this was one of the motivations for changing the class of triangulations
used in the path integral in CDT\footnote{It has recently 
been suggested that there might be a continuum limit of DT which belongs 
to the same universality class as the CDT theory described 
below \cite{laiho}.}. In the CDT formalism one sums over
geometries with a (proper) time foliation. 
In principle one starts out with space-times with a Lorentzian 
signature (contrary to the situation in DT) and the foliation is 
in proper (Lorentzian) time. However, each piecewise linear 
geometry used in the CDT path integral allows a 
rotation to Euclidean proper time. The set of Euclidean 
geometries we obtain in this way is a subset of the DT Euclidean 
geometries and this restriction 
seemingly cures some of the higher dimensional DT diseases, 
while in two dimensions
the relation between the restricted theory and the full DT theory 
has been worked out in detail: one obtains the CDT theory from 
the DT theory by integrating out all baby universes (which results
in a non-analytic mapping between the coupling constants 
of the two theories), and (somewhat
surprisingly) one can restore the DT theory from the CDT theory by the 
inverse mapping \cite{al,ackl}. 
Using four-simplices (which is the case having our attention in
this article) as building blocks   one can, for suitable choices of 
bare coupling constants, observe a four-dimensional 
(Euclidean) universe \cite{4dCDT}. 
For these choices of coupling constants the shape of 
the universe is consistent with an interpretation as an (Euclidean) de Sitter
space, at least as long as one looks at the scale factor \cite{planck}.
This is the region of coupling constants which will have our interest
(for other choices of the coupling constants one obtains more 
degenerate configurations \cite{phases}).

An interesting feature of the CDT model, rotated to 
Euclidean signature,  is that it possesses a transfer
matrix \cite{ajl2001,physrep}. 
At (discrete) time $t_n$ we have a spatial hypersurface with 
a spatial geometry characterized by a three-dimensional triangulation
$T_3(t_n)$. At time $t_{n+1}$ we have spatial geometry defined by another 
three-dimensional triangulation $T_{3}(t_{n+1})$. We assume for 
simplicity that the topology\footnote{$S^3$ is chosen for simplicity. 
We could have chosen any spatial topology. The important point in the 
assumption is that the topology is not allowed to change from one spatial 
hypersurface to the next.} of the spatial triangulations is that
of $S^3$. In CDT we sum over all four-dimensional triangulations 
of the ``slab'' between $t_n$ and $t_{n+1}$ compatible with the 
topology $S^3\times [0,1]$ and such that each four-simplex which 
``fills'' the slab has subsimplices which are (sub)simplices 
of $T_3(t_n)$ as well as $T_{3}(t_{n+1})$. 
This leads to 4 types of four-simplices in the slab:
type $(4, 1)$, $(3, 2)$, $(2, 3)$ and $(1, 4)$, where the numbers denote the 
number of vertices in $T_3(t_n)$ and $T_{3}(t_{n+1})$ respectively. The number
$N^{(4,1)}(t_n)$  of $(4, 1)$ simplices in the slab is equal to the 
number $N_3(t_n)$ of three-simplices in $T_3(t_n)$, and similarly 
the number of (1,4) simplices in the slab is equal to the number 
$N_3(t_{n+1})$ of three-simplices in $T_{3}(t_{n+1})$.
The transfer matrix $\cM$, i.e. the amplitude between the $T_3(t_n)$ and 
$T_{3}(t_{n+1})$, is now  given as the sum over all such triangulation,
\beql{ja2}
\la T_{3} (t_{n+1})| \cM | T_{3}(t_n)\ra =
\sum_{T_4} \frac{1}{C_{T_4}}\;\e^{-S[T_4]},
\eeq
where the summation is over all four-dimensional triangulations 
of a slab, with boundary triangulations $T_3(t_n)$ and $T_{3}(t_{n+1})$,
 $C_{T_4}$ is the order of the automorphism group of the triangulation $T_4$,
and where $S[T_4]$ is the Regge action of the four-dimensional 
triangulation of the slab. 

The transfer matrix is defined on the vector 
space spanned by the set $\cT_3$ of three-dimensional 
triangulations. This space is infinite dimensional and has the 
natural scalar product 
\beql{ja3}
\la T|T'\ra = \frac{1}{C_T} \del_{T,T'},~~~
\sum_T |T \ra C_T \la T| = \hat{I},~~~T,T' \in \cT_3
\eeq
where $C_T$ is the order of the automorphism group of the triangulation $T$.

The transition amplitude for a three-dimensional triangulation $T$ 
to develop into a three-dimensional triangulation $T'$ after $t_{tot}+1$
(integer) time steps is 
\beql{ja4}
\la T' | \cM^{t_{tot}+1} | T\ra = \sum_{\{T_i\}} \la T'|\cM|T_{t_{tot}}\ra 
C_{T_{t_{tot}}} 
\la T_{t_{tot}}|\cM |T_{t_{tot}-1}\ra \cdots C_{T_1} \la T_1|\cM|T\ra.
\eeq
We will define the partition function corresponding to $t_{tot}$ time steps 
as path integral with periodic boundary conditions after $t_{tot}$ time-steps:
\beql{ja5}
\cZ_{t_{tot}} = \tr \cM^{t_{tot}} = \sum_{T} C_{T} \la T | \cM^{t_{tot}}|T\ra.
\eeq
This is the partition function we have used in our 
computer simulations. The measurements performed so 
far, using Monte Carlo simulations, have been concentrated 
on the measurement of the scale factor, or more conveniently 
the three-volume $n_{t_i} \equiv N_3(t_i)$ at the spatial slice at time $t_i$,
as well as the correlation between the three-volumes at 
time $t_i$ and time $t_j$. These ``observables'' can be 
expressed using the transfer matrix $\cM$. The probability of 
measuring the spatial volume $n_{t_i}$ at time $t_i$ is given by 
\beql{ja6}
P^{t_{tot}}(n_{t_i}) = 
\frac{1}{\cZ_{t_{tot}}} \sum_{T\in \cT_3(n_{t_i})} C_T \la T| \cM^{t_{tot}}|T\ra
= \frac{ \tr {\trho}(n_{t_i}) \cM^{t_{tot}}}{\tr \cM^{t_{tot}}}.
\eeq
In \rf{ja6} $\cT_3(n_{t_i})$ denotes the subset  of 
three-dimensional triangulations where the number of three-simplices 
is $n_{t_i}$ and $\trho (n_{t_i})$ the projection operator on the 
subspace spanned by these triangulations:
\beql{ja7}
\trho(n) = \sum_{T \in \cT_3(n)} |T\ra C_T \la T|, \quad \trho(n)^2 = \trho(n).
\eeq

Similarly the correlator between $n_{t_1}$ and $n_{t_2}$, separated
by $\Del t = t_2-t_1$ is given by 
\bea
P^{t_{tot}}(n_{t_1},n_{t_2}) &=& \frac{1}{\cZ_{t_{tot}}} 
\sum_{\substack{T_1\in \cT_3(n_{t_1})\\
T_2 \in \cT_3(n_{t_2})}} C_{T_1} 
\la T_1| \cM^{t_{tot}-\Del t} |T_2\ra C_{T_2} \la T_2| 
\cM^{\Del t} |T_1\ra \nonumber \\
&=& \frac{ \tr \trho(n_{t_1}) \cM^{t_{tot}-\Del t} \trho(n_{t_2}) 
\cM^{\Del t}}{\tr \cM^{t_{tot}}}, \label{ja8}
\eea 
and this expression can clearly be generalized to multi-correlators.

As mentioned in the Introduction it was possible to explain the 
observed distributions 
$P^{t_{tot}}(n_{t_1})$ and $P^{t_{tot}}(n_{t_1},n_{t_2})$
using a simple minisuperspace model. How does the concept of 
a minisuperspace labeled by states $n_{t_i}$, $i=1,\ldots,t_{tot}$
relate to the transfer matrix $\cM$ which is defined on the much 
larger space spanned by the vectors $\cT$? Let us define 
the ``effective'' transfer matrix $M$ by 
\beql{ja9a}
|n\ra\la n|M|m\ra\la m| = \sum_{\substack{T_n \in \cT_3(n)\\ T_m \in\cT_3(m)}} 
\;|T_n\ra C_{T_n}\la T_n | \cM|T_m\ra C_{T_m}\la T_m|.
\eeq
$\la n|M|m\ra$ represents the 
average of the matrix elements $\la T_n | \cM|T_m\ra$ like
\beql{ja9}
\la n|M|m\ra = \la \cM \ra_{n,m} := \frac{1}{{\cN_n \cN_m}} 
\sum_{\substack{T_n \in \cT_3(n)\\ T_m \in\cT_3(m)}} \sqrt{C_{T_n}C_{T_m}} 
\; \la T_n | \cM|T_m\ra,      
\eeq
where $\cN_n$ denotes the cardinality of $\cT_n$. $\cN_n$ grows
exponentially with $n$. 

In \rf{ja9a} and \rf{ja9} it is misleading to think of the ``state'' $|n\ra$ as
the (suitably) normalized sum of the $\cN_n$ vectors $|T_n\ra$ (although
by doing so one would of course obtain the correct expectation 
value $\la \cM \ra_{n,m}$ using such vectors).
Such a vector would again be a single vector located in the 
$\cN_n$-dimensional space spanned by the $|T_n\ra$'s.
It is more appropriate to think of the ``state'' associated with $n$ 
as  arising from a uniform {\it probability distribution} of states
$|T_n \ra$  and in this way to think of $\trho(n)$ as the associated
{\it density operator}. However, once we have reduced our 
consideration to the matrix  $\la n| M|m\ra$ we are of course 
free to find eigenvectors for this matrix and expand them in 
the abstract basis $|n\ra$, and we will indeed do that.

The statement that we can use the matrix $\la n|M|m\ra$ as an 
effective transfer matrix is the statement that the standard deviation
of the $\cN_n\cN_m$ numbers $\la T_n | \cM | T_m\ra $ is sufficiently small.
In fact the difference between $\tr \cM^2$ and $\tr M^2$ can exactly 
be expressed as a sum over deviations squared for each $n,m$: 
\beql{ja10} 
\tr \cM^2 - \tr M^2 = \sum_{n,m}
\sum_{\substack{T_n\in \cT_3(n)\\ T_m\in \cT_3(m)}}  \left( \sqrt{C_{T_n}C_{T_m}} 
  \la T_n| \cM|T_m\ra - \la \cM \ra_{n,m}\right)^2
\eeq  

In the following we will assume that we can work with an 
effective transfer matrix $\la n|M|m\ra$. Eq.\ \rf{ja9a} is an attempt
to {\it define} this effective transfer matrix from first principles
and in principle one can check by computer simulations 
if it is a good approximation. We will here take the pragmatic 
attitude to assume there exists such an effective transfer 
matrix and use it to analyze the computer generated data.  
The consistency of this analysis is indirectly evidence 
that an object like $\la n|M|m\ra$ provides a good approximation
of our data. Thus we will use the ``effective'' version of 
\rf{ja6}-\rf{ja8}:
\beql{ja11}
\rho(n) = | n\ra\la n|,
\eeq
\beql{ja12}
P^{t_{tot}}(n_{t_i}) = 
\frac{ \tr {\rho}(n_{t_i}) M^{t_{tot}}}{\tr M^{t_{tot}}}.
\eeq
\beql{ja13}
P^{t_{tot}}(n_{t_1},n_{t_2}) =
\frac{ \tr \rho(n_{t_1}) M^{t_{tot}-\Del t} \rho(n_{t_2}) 
M^{\Del t}}{\tr M^{t_{tot}}},
\eeq 
where $\rho(n)$ should be distinguished from $\trho(n)$. 

In particular we can measure the matrix elements 
$\la n|M|m\ra$ up to a normalization by considering $t_{tot} = 2$. We have:
\beql{ja13a}
P^{(2)}(n_1,n_2) = \frac{\la n_1|M|n_2\ra \la n_2 |M|n_1\ra}{\tr M^2}
\eeq 
This method requires a major change in our general computer program which assumes $t_{tot} \geq 3$.
We updated it but we were not completely convinced that the new version is stable
although it gave exactly the same results as the method we finally used.
For $t_{tot} = 3, 4$ we get:
\bea\label{ja13b}
P^{(3)}(n_1,n_2) &=& \frac{\la n_1|M|n_2\ra \la n_2 |M^2|n_1\ra}{\tr M^3}\\
P^{(4)}(n_1,n_3) &=& \frac{\la n_1|M^2|n_3\ra \la n_3 |M^2|n_1\ra}{\tr M^4}
\label{ja13c}
\eea
From the measurements of $P^{(3)}(n_1,n_2)$ and $P^{(4)}(n_1,n_3)$
we can determine the matrix elements $\la n|M|m\ra$ up to a normalization:
\beql{ja13d}
\la n|M|m\ra = C \; \frac{P^{(3)} (n_1=n,n_2=m)}{\sqrt{P^{(4)}(n_1=n,n_3=m)}}
\eeq

There is nothing magic about the above choice. One could have chosen 
$t_{tot}=4$ and $t_{tot} = 6$ and formed the combinations
\bea\label{ja13e}
P^{(4)}(n_1,n_2) &=& \frac{\la n_1|M|n_2\ra \la n_2 |M^3|n_1\ra}{\tr M^4}\\
P^{(6)}(n_1,n_4) &=& \frac{\la n_1|M^3|n_4\ra \la n_4 |M^3|n_1\ra}{\tr M^6}
\label{ja13f}
\eea
from which one can again extract $\la n|M|m\ra$ like in \rf{ja13d}. 
We have indeed checked that measurements of $P^{(4)}(n_1,n_2)$ and
$P^{(6)}(n_1,n_4)$ lead to the same $M$ matrix as extracted from 
measurements of $P^{(3)}(n_1,n_2)$ and $P^{(4)}(n_1,n_3)$, up to 
a normalization.

In earlier work we have shown \cite{planck,semiclassic} that the following
minisuperspace action\footnote{In fact we used slightly different form of the potential terms: $ \mu 
\left(\frac{n_t+n_{t+1}}{2}\right)^{1/3}
-\lam \frac{n_t+n_{t+1}}{2} \Rightarrow  \mu 
n_t^{1/3}
-\lam n_t$. This parametrization was more convenient to extract the parameters of the action from the measured covariance matrix of volume fluctuations. In this article we implement a modified form (\ref{ja14}) which better fits our data.}:
\beql{ja14}
S[\{n_t\}] =   \sum_t \frac{1}{\Gamma} \left[ \frac{(n_{t+1}-n_t)^2}{n_t+n_{t+1}} + \mu 
\left(\frac{n_t+n_{t+1}}{2}\right)^{1/3}
-\lam \frac{n_t+n_{t+1}}{2} \right],
\eeq
describes well the measured $\la n_t \ra$ and the fluctuations 
$\la n_t n_{t'}\ra - \la n_t \ra\la n_{t'} \ra$ in 
the bulk where $n_t$ is large. 
The effective action \rf{ja14} suggests  that  the  effective transfer
matrix
\beql{ja15}
\la n | M | m\ra \propto  e^{ 
- \frac{1}{\Gamma} \left[ \frac{(n - m)^2}{n + m} + \mu 
\left(\frac{n + m}{2}\right)^{1/3}
-\lam \frac{n + m}{2} \right]}
\eeq
is a good approximation in the bulk. 
We will in the following try to determine the transfer matrix 
from the data, also in the range where $n_t$ is not necessarily
large and we will try to improve the expression \rf{ja15}.

\section{How to perform the computer simulations}

The simplest version of the discretized CDT theory  has 
three parameters, two related to the cosmological constant and  
the gravitational constant, and an additional parameter which controls 
the  asymmetry between the edge lengths in the spatial and time directions.
This latter parameter seems not to be a genuine coupling constant
since it just labels the different length assignment of spatial and 
time-like links. The action used is still the 
Einstein-Hilbert action (as formulated by Regge for 
piecewise linear geometries), adjusted for this asymmetry. 
However, because we study the theory in a truly
non-perturbative region of coupling constant space the effective
action is determined by a competition between the classical 
action used and a contribution coming from the measure term. 
The contribution from the measure term is ``entropic'' in nature:
it counts the number of configurations with the same action and 
is thus independent of the other parameters. Effectively this 
promotes the asymmetry parameter to a genuine coupling constant
(see \cite{physrep} for a detailed discussion).    
  
In the numerical simulations the topology of the manifold is assumed to be 
$S^3 \times S^1$ with periodic boundary conditions in the (Euclidean) time,
as mentioned above.  The four-simplices used to 
construct the simplicial manifolds of CDT are characterized 
by their position in spatial and time directions.
As also mentioned above we have  four types of  four-simplices: 
$(4,1)$-simplices, with four vertices 
at time $t$ and one vertex at $t+1$, $(3,2)$-simplices 
with three simplices at $t$ and two
at $t+1$ and the ``time-reversed'' $(1,4)$-simplices and 
$(2,3)$-simplices. All simplices of a particular type are identical.

The discretized (Regge) Einstein-Hilbert action becomes extremely simple
because we are essentially only  using the two kinds of building blocks
to construct the four-dimensional triangulation $T$   
\cite{ajl2001,physrep}:
\begin{equation}
S_R[T] =  -(\kappa_0+6\Delta) N_0 + 
\kappa_4 \left( N^{(4,1)}+ N^{(3,2)}\right)+
\Delta \left(  2 N^{(4,1)}+ N^{(3,2)}\right)
\label{Sdisc}
\end{equation}
where  $N_0$ is the total number of vertices in the triangulation,  
$N^{(4,1)}$ the total number of type ${(4,1)}$ plus ${(1,4)}$ simplices 
and $ N^{(3,2)}$ the  total number of simplices of 
type ${(3,2)}$ plus ${(2,3)}$. 
$\kappa_0$, $\kappa_4$ and $\Delta$ 
are the (bare) dimensionless coupling constants 
obtained by the discretization of the continuous action (\ref{SHE}).
$\kp_0$ is proportional to the inverse bare gravitational constant,
$\kp_4$ related to the cosmological constant while $\Del$ is related
to the asymmetry between the spatial and time-like links. $\Del =0$
corresponds to spatial and time-like links having the same length.   
An additional geometric parameter is the length $t_{tot}$ 
of the periodic time axis.

\begin{figure}[t]
\centering
\scalebox{0.8}{\includegraphics{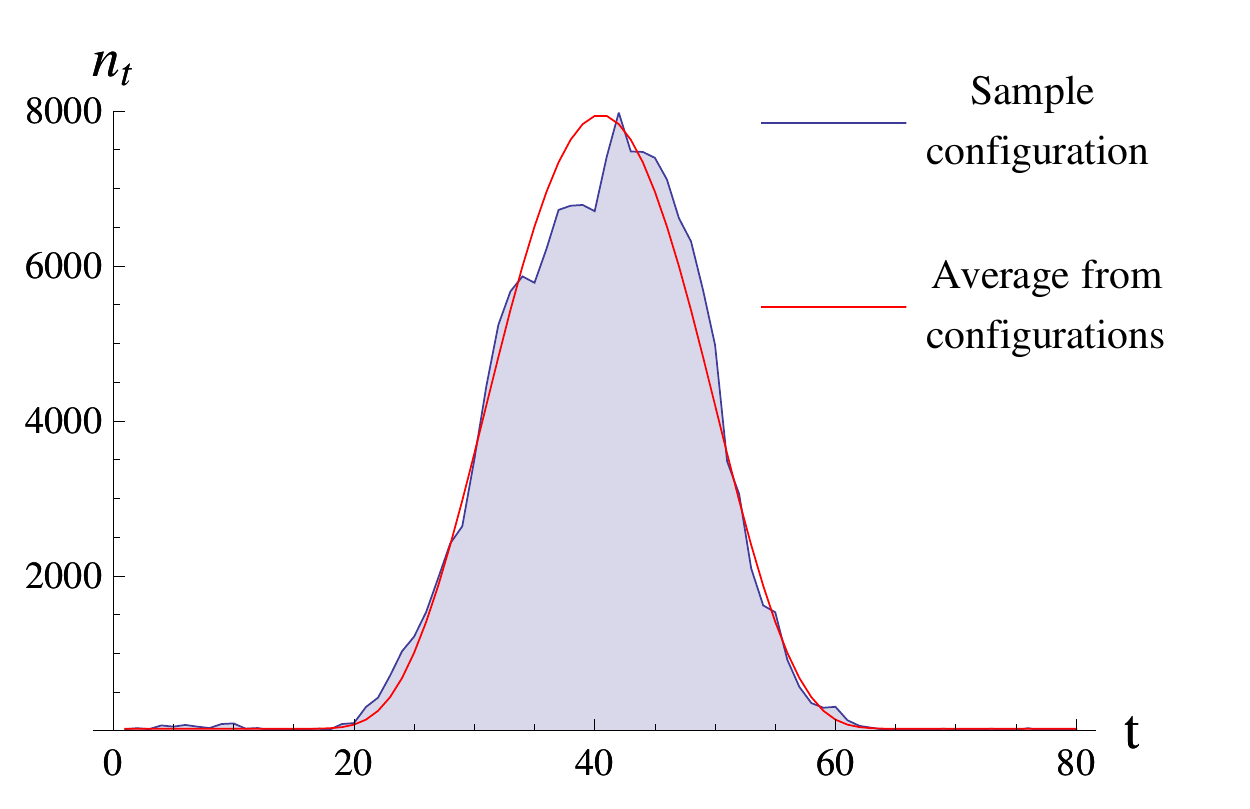}}
\caption{The distribution $n_t$ of the three-volume, i.e.\ the 
number of $(4,1)$-simplices,  
when we are in the de Sitter phase. The data is 
obtained for $\kappa_0 = 2.2$, $\Delta =0.6$. A single (typical) configuration
as it appears in the path integral  is shown by the blue line, and  
the  average distribution is indicated by the red line.}
\label{Fig01}
\end{figure}

The partition function 
\begin{equation}\label{ja18}
{\cal Z} = \sum_{T\in {\cT_4}} e^{-S_R[T]}
\end{equation}
has a critical value $\kp_4^{crit}(\kp_0,\Del)$, depending
on $\kp_0$ and $\Del$, such that $\cZ$ is divergent for 
$\kp_4 < \kp_4^{crit}$. The existence of this critical value is 
reflecting the fact that the number of triangulations with 
a fixed number of four-simplices $N_4$ grows exponentially with $N_4$.
In principle we want to fine tune $\kp_4$ to this critical value 
since we really want a limit where $N_4\to \infty$. In practice the
simulations have so far 
been carried out by keeping $N_4$ (or $N^{(4,1)}$) fixed. In this way we 
have been trading the coupling constant $\kp_4$ with $N_4$ and the 
partition function $\cZ(N_4)$ is related to $\cZ(\kp_4)$ by a Laplace 
transformation:
\beql{jan20}
\cZ(\kp_4) = \sum_{N_4} e^{-\kp_4 N_4} \cZ(N_4).
\eeq
The phase diagram now depends on $\kp_0$ and $\Del$ and we refer 
to \cite{physrep,phases} for a detailed discussion. Here we will be 
working in the interesting, so-called de Sitter phase where 
we, for a given (large) $N_4$ and sufficient large $t_{tot}$, 
observe a (Euclidean) de Sitter universe, i.e. a four-sphere 
where the temporal extension is  proportional to $N_4^{1/4}$
while the rest of the time extension (assuming $t_{tot}$ is large enough
compared to $N_4^{1/4}$) is a {\it stalk} of almost no spatial 
extension. Presumably this stalk only exists because our computer 
algorithm does not allow the spatial extension to shrink to zero. 
In Fig.\ \ref{Fig01} we have shown a typical situation with $t_{tot}=80$,
$N^{(4,1)} = 160 000$ and we observe a bulk region (the ``blob'', approximately from 
$t=20$ to $t=60$) where $n_t$, the three-volume, 
i.e.\ the number of tetrahedra at the time-slice $t$, 
is large,  and the rest is the stalk region where $n_t$ is very small.
Fig.\ \ref{Fig01} shows both the average over many configurations and
a typical configuration which appears in the path integral. When taking 
the average over many configuration we align the center of mass of the 
blobs (see \cite{planck} for a detailed discussion). 

\begin{figure}[t]
\centering
\includegraphics[width=0.48\textwidth]{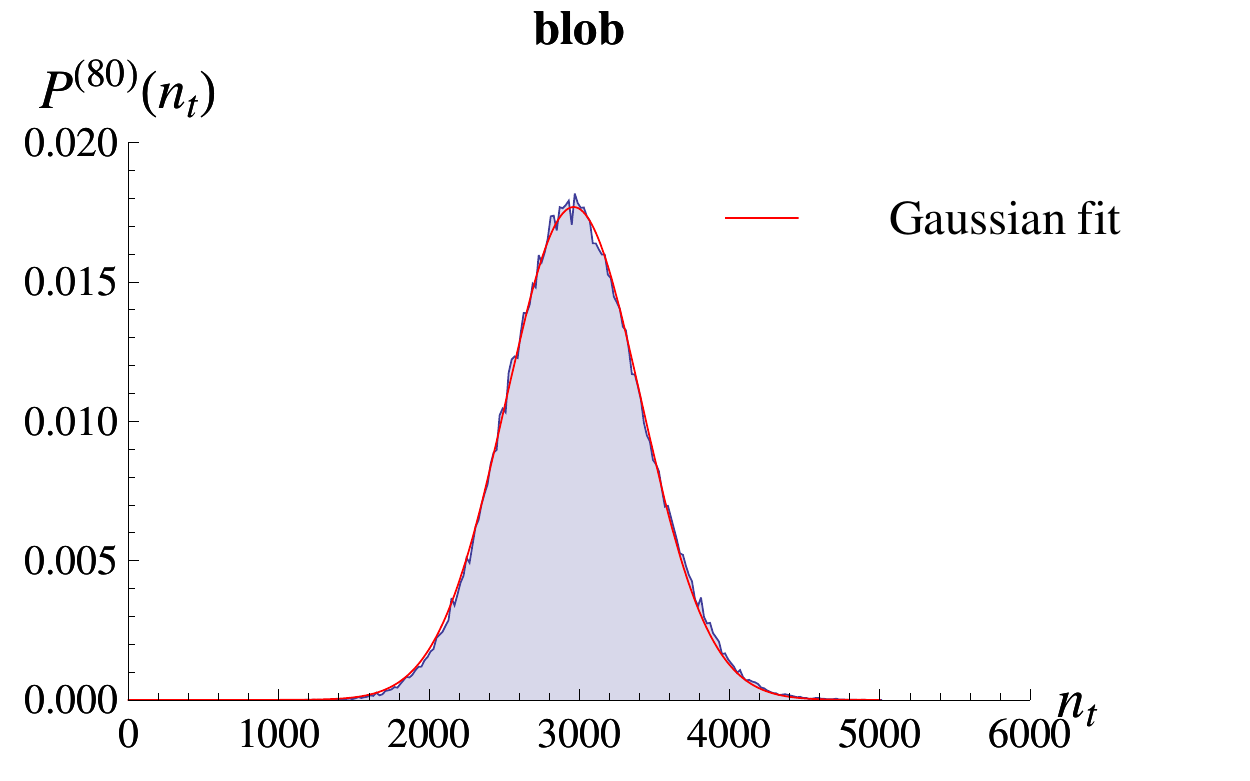}
\includegraphics[width=0.44\textwidth]{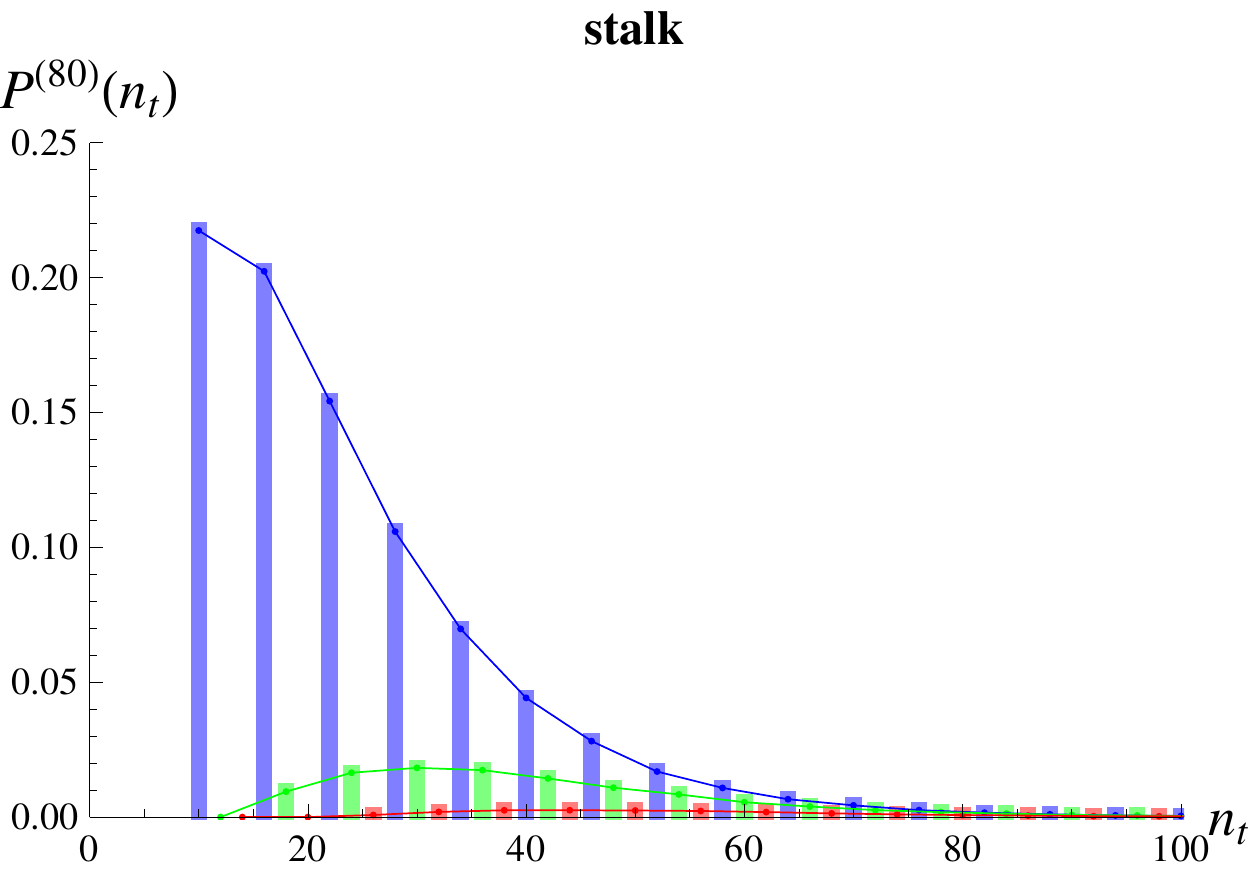}
\caption{Left figure:  Probability distribution of the volume $n_t$ 
in {\em the blob} ($t = 29$ in Fig.\ \ref{Fig01}), 
Right figure: Probability distribution in {\em the stalk} 
($1\leq t \leq 17$ in Fig.\ \ref{Fig01}).} 
\label{Fig02}
\end{figure}

We can also measure the probability distribution $P^{t_{tot}} (n_t)$ of 
$n_t$ in the blob for a given $t$ in Fig. \ref{Fig01}. 
It is shown in Fig.\ \ref{Fig02} (left figure). It
is well approximated by a Gaussian distribution around the mean value
$\la n_t\ra$. This is in contrast to the situation in the stalk 
where the probability distribution splits in three families \cite{we}, as 
shown on the right part of Fig.\ \ref{Fig02}.  

\begin{figure}[t]
\begin{center}
\includegraphics[width=0.8\textwidth]{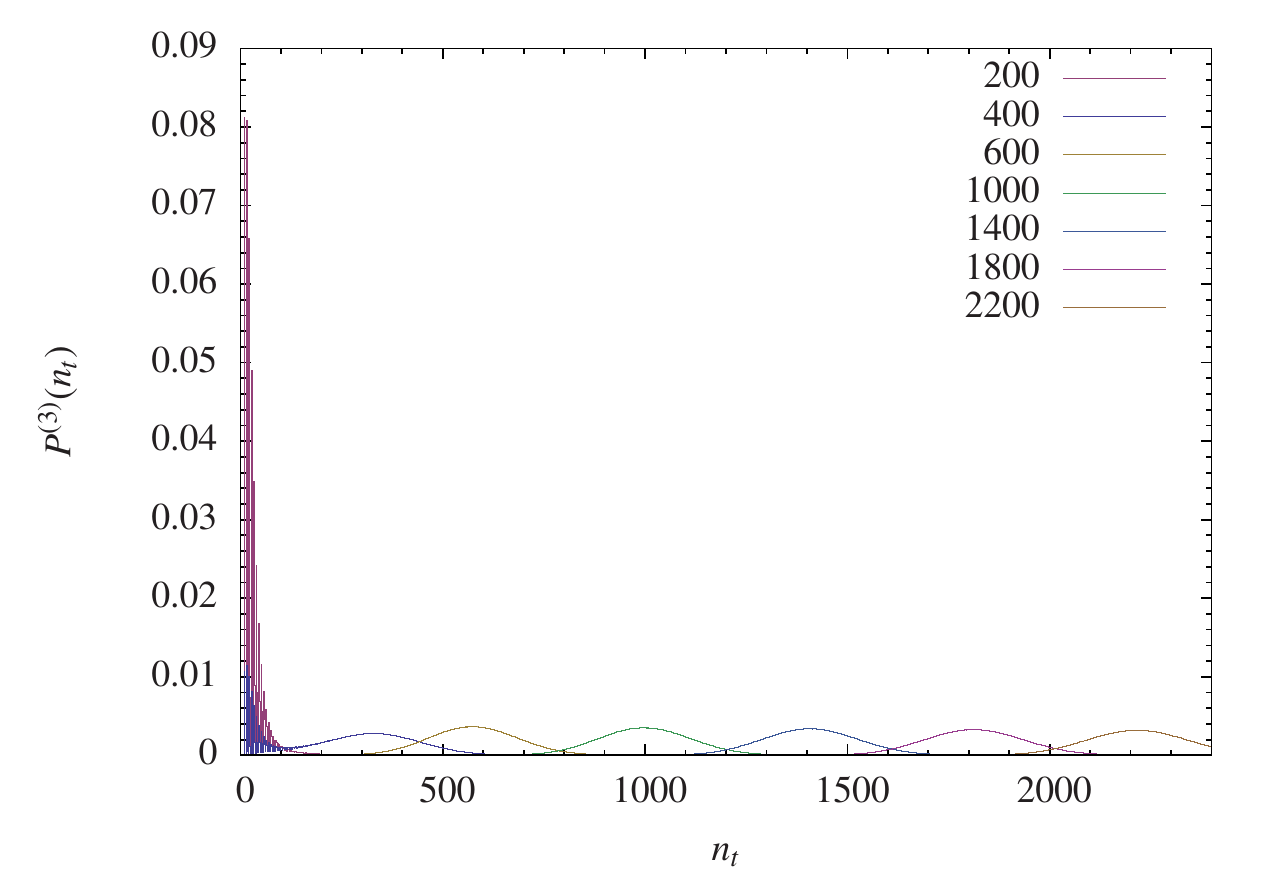}
\end{center}
\caption{
Probability distribution of the volume $n_t$,
for various $n_{vol} = 200, 400, 600, 1000, 1400, 1800$ and $2200$ 
(from left to right).
For all ranges, the simulations were performed 
with $\kappa_4 = 0.322$ and $\epsilon = 0.00002$.
}
\label{FigHistSeries}
\end{figure}

Figs.\ \ref{Fig01} and \ref{Fig02} are based on computer simulations
of the type mentioned above: $N^{(4,1)}$ is kept fixed. Technically this 
has been done by adding a term $\ep (N^{(4,1)}- \bar{N}^{(4,1)})^2$ to the action,
$\ep$ being a suitably small parameter:
\beql{ja19}
S_R \to S_R +\ep (N^{(4,1)}- \bar{N}^{(4,1)})^2.
\eeq
This term ensures that $N^{(4,1)}$ is going to fluctuate not too far 
from $\bar{N}^{(4,1)}$. The precise value $\la N^{(4,1)} \ra$ depends
on the choice of $\kp_4$.
 We now want to study the transfer matrix. However, the structure
of the transfer matrix is incompatible with a global constraint of 
this type, so we have to change the updating procedure. We have done
this in two different ways. The first way is  to drop the 
constraint term $\ep (N^{(4,1)}- \bar{N}^{(4,1)})^2$ and only use the discretized 
Einstein-Hilbert action \rf{Sdisc}. The way to obtain an average 
$\la N^{(4,1)}\ra$ is to fine tune $\kp_4$ to $\kp_4^{crit}$. The closer 
$\kp_4$ is to the critical value the larger $\la N^{(4,1)}\ra$. In practice
this fine tuning can be difficult and the larger the system
the more difficult the fine tuning. Thus we can and will only 
use it for small systems. For larger systems we apply a different 
strategy which also constrains the value of $N^{(4,1)}$, but which is compatible 
with the transfer matrix structure: we change the global constraint
imposed on $N^{(4,1)}$ in \rf{ja19} to a local constraint in $t$:  
\beql{ja21}
S_R \to \tS_R= S_R + \ep \sum_{t=1}^{t_{tot}} (n_t -n_{vol})^2.
\eeq
Of course this constraint will drastically change the profile 
$\la n_t \ra$, since $n_t$ will now fluctuate around $n_{vol}$.
Thus we will have different transfer matrices $\tilde{\cM}$ and $\tM$,
and different probability distributions $\tP(n_1,n_2,\ldots)$.
The new probability distribution for $n_t$ is shown  for various $n_{vol}$ 
in Fig. \ref{FigHistSeries}.

However, we can reconstruct some of the probability distributions
associated with the action $S_R$ if we know it (i.e.\ measure it) 
for the action $\tS_R$. The probability
for measuring $(n_1,n_2,\ldots,n_{t_{tot}})$ is given by   
\begin{equation}
\tilde{P}^{t_{tot}}(n_1, n_2, \dots, n_{t_{tot}}) 
= \frac{\langle n_1 | \tilde{M} | n_2 \rangle \langle n_2 | 
\tilde{M} | n_3 \rangle \cdots 
\langle n_{t_{tot}} | \tilde{M} | n_1 \rangle}{\tr \tilde{M}^{t_{t_{tot}}}},\\ 
	\label{EqTildePTM}
\end{equation}
and is directly related to the distribution without the volume fixing term,
\begin{equation}
	\tilde{P}^{t_{tot}}(n_1, n_2, \dots, n_{t_{tot}}) 
\propto P^{t_{tot}}(n_1, n_2, \dots, n_{t_{tot}}) 
e^{-\epsilon (n_1 - n_{vol})^2} \cdots 
e^{-\epsilon (n_{t_{tot}} - n_{vol})^2}.\\
	\label{EqPTildeP}
\end{equation}
We calculate the transfer matrix $\tilde{M}$ in the same way as $M$:
\begin{equation}
	\langle n | \tilde M | m \rangle = 
\frac{\tilde{P}^{(3)} (n_1=n, n_2=m)}{\sqrt{\tilde{P}^{(4)}(n_1=n, n_3=m)}}.
\end{equation}
To calculate the original transfer matrix $M$ 
we have to cancel the volume fixing term, which is easily done:
From equations (\ref{ja21}), (\ref{EqTildePTM}) and (\ref{EqPTildeP}) we obtain
\begin{equation}\label{ja22}
	\langle n | M | m \rangle = 
e^{\frac{1}{2}\epsilon (n - n_{vol})^2} 
\langle n | \tilde M | m \rangle e^{\frac{1}{2}\epsilon (m - n_{vol})^2}.
\end{equation}

For each choice of $n_{vol}$ we observe $n_t$ with some 
approximate Gaussian distribution centered around $n_{vol}$,
where the width depends on our choice of $\ep$,
and we use the associated probabilities to construct $\la 
n|M|m\ra$, as described above. To reconstruct the matrix in 
a larger region of the $n$-space we have to merge data from  
different $n_{vol}$ regions. Since the matrix is determined 
only up to a normalization, the way to do this is to make 
sure there are regions of overlap between the $n_t$ distributions
and in these regions choose a suitable calibration procedure
such that we can merge the data. We will later describe how 
this is explicitly done. Needless to say the procedure we 
are employing here is a kind of multi-canonical Monte Carlo 
method (see \cite{berg} for a review).

\section{The effective action at large three-volumes}

\begin{figure}[t] 
\begin{center}
\includegraphics[width=0.4\textwidth]{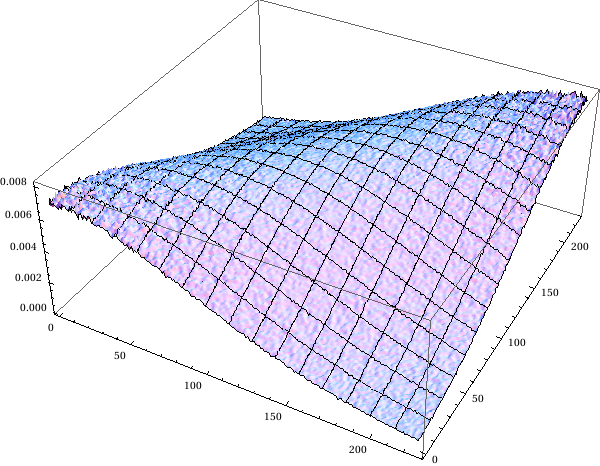}
\includegraphics[width=0.4\textwidth]{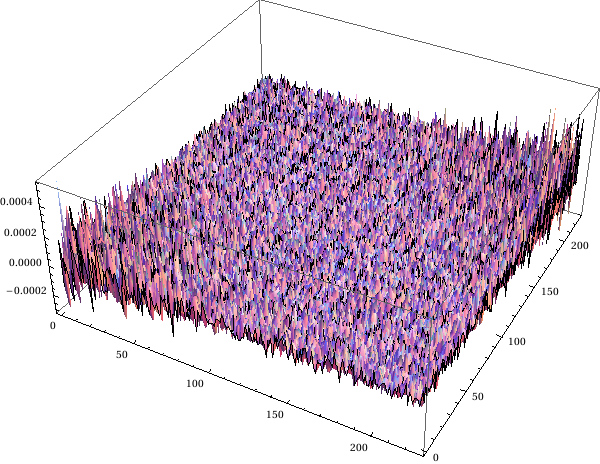}
\end{center}
\caption{
The left figure: The transfer matrix $M$ for range $1200  < n_t < 1600$. The matrix is optically indistinguishable from the  fitted theoretical transfer matrix $M^{(th)}$.
The right figure: The difference between $M$ and $M^{(th)}$  
disappears in the numerical noise. The measurements were performed 
for $\kp_0 = 2.2, \Delta = 0.6, \kp_4 = 0.3220$ and $\epsilon = 0.00002$.
}
\label{FigTMXL}
\end{figure}

We can measure the transfer matrix $\la n| M |m\ra$ for large
$n,m$ as described above. As already noted it is well approximated
by the matrix $M^{(th)}$:
\begin{equation}
	\langle n | M^{(th)} | m \rangle = \cN e^{-L_{eff}(n, m)},
	\label{EqMth}
\end{equation}
where the effective Lagrangian is 
\begin{equation}
 L_{eff} (n, m) = \frac{1}{\Gamma} 
\left[ \frac{(n - m)^2}{n + m -2 n_0} + 
\mu \left( \frac{n + m}{2} \right)^{1/3} - 
\lambda \left( \frac{n + m}{2} \right) \right].
	\label{Eqseff}
\end{equation}
We now ask how well? We make a best fit of the parameters
$\Gamma, \mu, \lambda$ and $\cN$ (fixing $n_0=0$).  
The measured $M$, $M^{(th)}$ from \rf{EqMth} as well as their difference,
are shown in Fig. \ref{FigTMXL} for $n_{vol} = 1400$ 
and for the  $n_t$ range $1200  < n_t < 1600$.

The values of parameters $\Gamma, \mu$ and $\lambda$ for 
different values of $n_{vol}$, obtained from 
the best fits of the  matrix $M^{(th)}$ (\ref{EqMth})
to the measured matrix $M$ are presented in Table \ref{table1}.
Again $n_0$ is chosen to be zero.

\begin{table}[ht]
	\begin{center}
		\begin{tabular}{|c|c|c|c|c|}
\hline
$n_{vol}$ & $n_t$ range	& $\Gamma$ & $\mu$ & $\lambda$ \\
\hline
\hline
$600$	& $ 400 - 820$	& $25.71 \pm 0.01$	& $18 \pm 1$	& $0.05 \pm 0.01$ \\ \hline
$1000$	& $ 780 - 1220$	& $26.00 \pm 0.01$	& $17 \pm 1$	& $0.05 \pm 0.01$ \\ \hline
$1400$	& $1180 - 1630$	& $26.10 \pm 0.01$	& $13 \pm 1$	& $0.04 \pm 0.01$ \\ \hline
$1800$	& $1580 - 2040$	& $26.08 \pm 0.01$	& $26 \pm 1$	& $0.07 \pm 0.01$ \\ \hline
$2200$	& $1980 - 2440$	& $26.05 \pm 0.01$	& $19 \pm 2$	& $0.05 \pm 0.01$ \\ \hline
		\end{tabular}
	\end{center}
	\caption{The values of 
$\Gamma, \mu$ and $\lambda$  for different $n_{vol}$, obtained from best fits
of $M^{(th)}$ to the measured $M$.}
	\label{table1}
\end{table}

\subsection{The kinetic term}

To get a better estimation of the parameters associated with
the effective action (\ref{Eqseff}),
we first try to fit only to the 
parameters of the kinetic term which is by far the 
dominating term from a numerical point of view. 
We do that by keeping the sum of the entries, i.e. $n+m$, fixed
such that the potential term is not changing. In this way we can 
try to determine $\Gamma$ and even $n_0$ which we had put to zero 
in the fits mentioned above in order not to have too many fit-parameters.
\begin{figure}
\begin{center}
\includegraphics[width=0.8\textwidth]{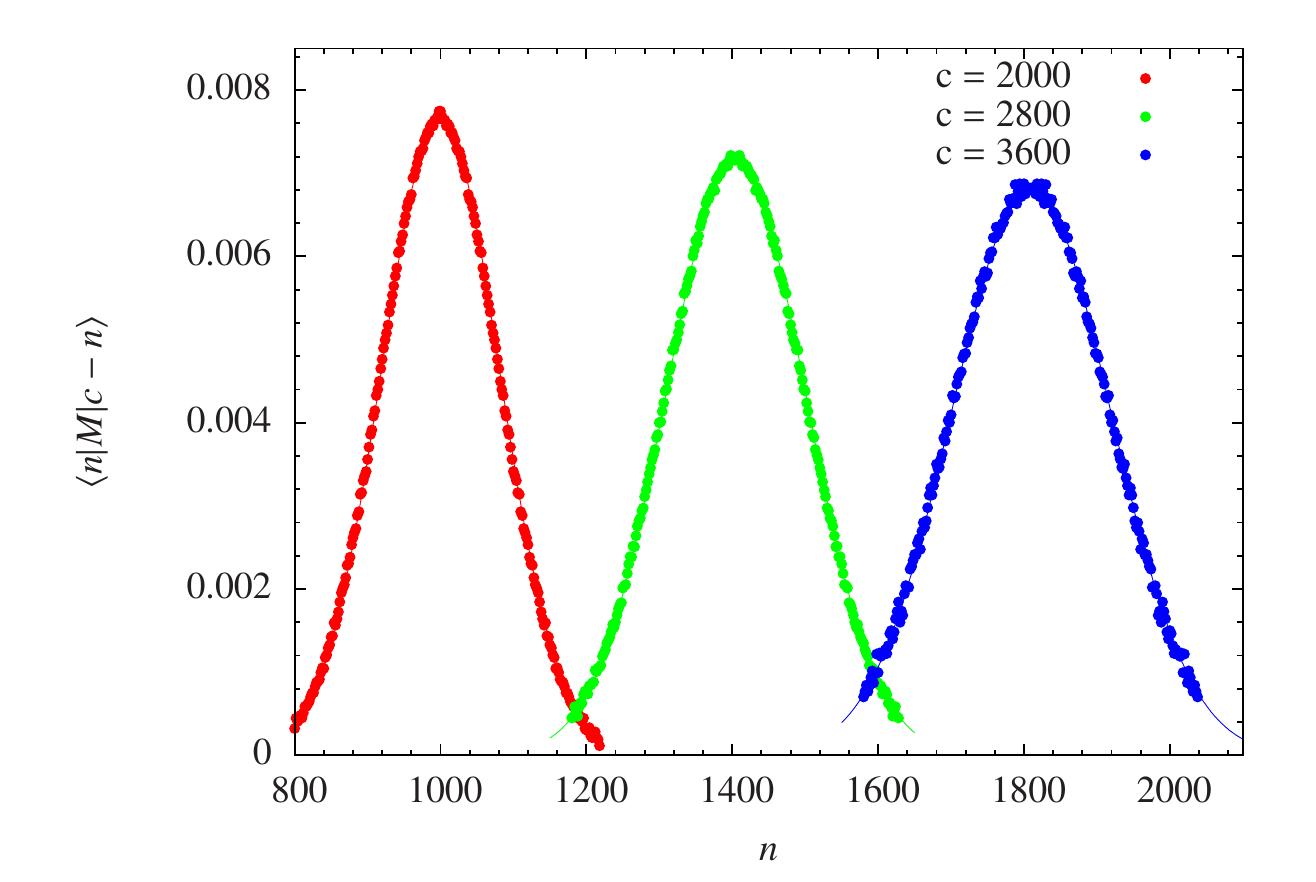}
\end{center}
\caption{
$\langle n | M | c - n \rangle$ plotted as a 
function of $n$ for $c = 2000,2800$ and $3600$ (dots).
Gaussian fits are drawn with a line.}
\label{FigAntidiagonals}
\end{figure}
The matrix elements  for constant $n + m = c$
show the expected Gaussian dependence on $n$ 
(see Fig. \ref{FigAntidiagonals}):
\beql{ja30}
\langle n | M | m \rangle = \langle n | M | c - n \rangle = 
\mathcal{N}(c) \exp \left[- \frac{(2 n - c)^2}{\Gamma \cdot (c - 2 n_0)} \right], 
\eeq
where the terms in the effective action which only depend on $c$ are 
included in the normalization.

\begin{figure}
\begin{center}
\includegraphics[width=0.8\textwidth]{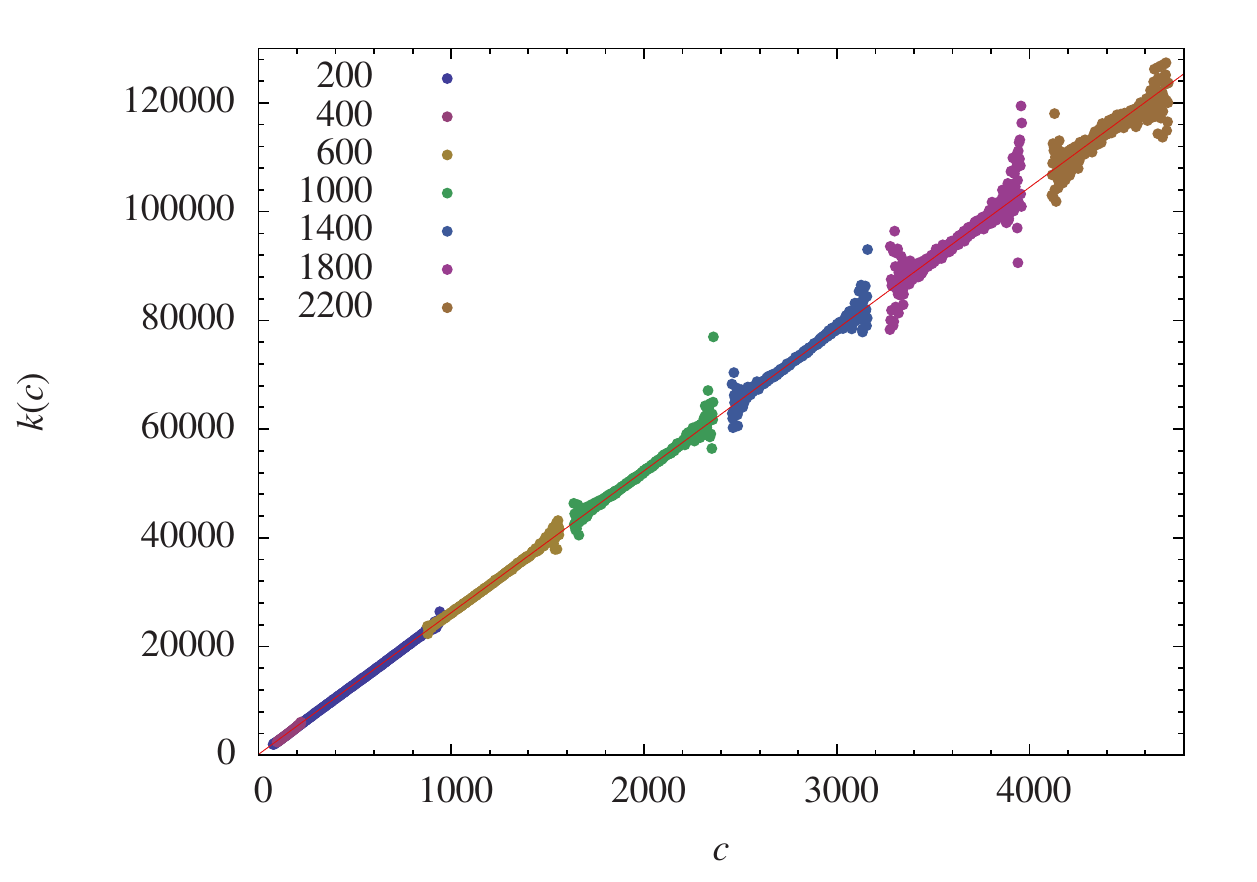}
\end{center}
\caption{
The coefficient $k(c)$ in the kinetic term 
as a function of $c = n + m$ 
(different colors denote different ranges),
and a linear fit $k(n + m) = \Gamma \cdot  (n + m - 2 n_0)$ (red line).
}
\label{FigKinetic}
\end{figure}

We expect the denominator of the kinetic term $k(c)$ to behave like 
$k(n + m) = \Gamma \cdot (n + m - 2 n_0)$.
As shown on Fig. \ref{FigKinetic} 
this is indeed true and the parameter $\Gamma$ is common for all ranges.
Fig. \ref{FigKinetic} presents measured 
coefficients $k(c)$ for various $c$'s and ranges of $n_t$
denoted by distinct colors together with a linear fit.
The best linear fit gives $\Gamma = 26.07$ and $n_0 = - 3$,
which is consistent with results obtained for separate ranges of $n_t$.

\subsection{The potential term}

\begin{figure}
\begin{center}
\includegraphics[width=0.8\textwidth]{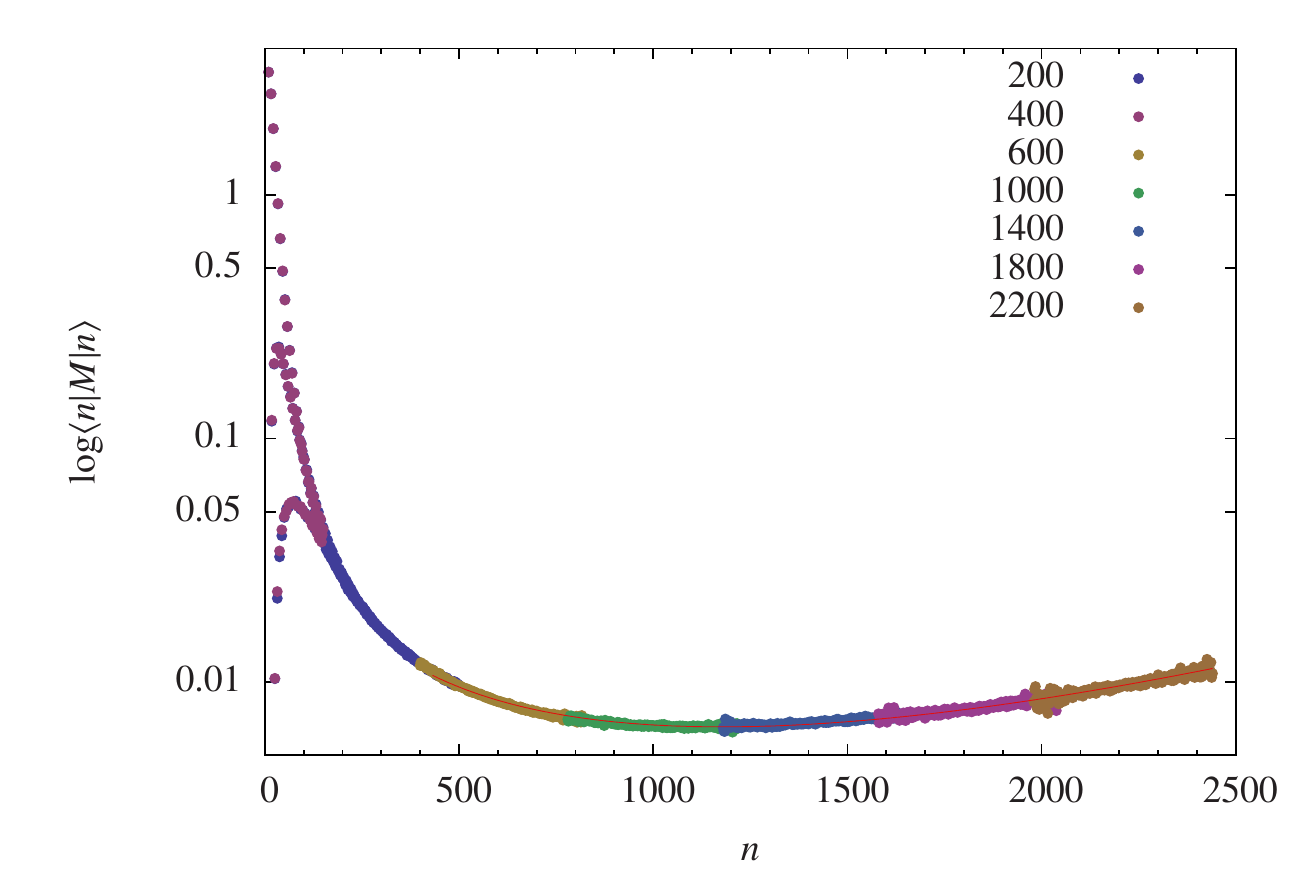}
\end{center}
\caption{
$\log \langle n | M | n \rangle$ of the scaled transfer matrix 
(dots, different colors denote different ranges)
compared with the fit of the potential term $-L_{eff}$ (red line,
which stops at $n=400$).
}
\label{FigPotential}
\end{figure}

The potential part of the effective Lagrangian 
may be extracted from the diagonal elements of the transfer matrix
\begin{equation}
 L_{eff}(n, n) = - \log \langle n | M | n \rangle + c(n_{vol}) = 
\frac{1}{\Gamma} \left( \mu n^{1/3} - \lambda n \right).
 \label{EqSeffPot}
\end{equation}
However, because of different normalizations of the transfer matrices 
for different ranges (hence the dependence of the 
constant $c(n_{vol})$ on $n_{vol}$), 
the fit of $L_{eff}(n, n)$ to the transfer 
matrix data cannot be performed  in a straightforward way. 
The transfer matrices have first to be  merged properly 
via a scaling procedure, i.e. by adjusting the $c$ constant 
in (\ref{EqSeffPot}). This is done in the following way.
For example, for $n_{vol} = 1400$ the range of spatial volumes 
for which we measured the transfer matrix
is $n_t = 1180 \dots 1630$, while for $n_{vol} = 1800$ 
the range is $n_t = 1580 \dots 2040$.
Thus, there is a non-vanishing intersection $n_t = 1580 \dots 1630$
for which elements of both matrices were measured.
We scale the second matrix, so that the mean value of the 
diagonal elements on the intersecting region is equal for both matrices.
After applying this procedure for successive ranges,
we finally get scaled transfer matrices which can be merged.
The result of such merging is shown on Fig.\ \ref{FigPotential},
which shows the diagonal elements of the scaled transfer matrices 
together with a fit of form (\ref{EqSeffPot}), where we took $\Gamma = 26.1$.

\subsection{A global effective action fit}

\begin{table}
	\begin{center}
		\begin{tabular}{|c|c|c|c|c|}
\hline
 Method 			& $\Gamma$ 				& $n_0$ 		& $\mu$ 			& $\lambda$ \\
\hline
\hline
 Cross-diagonals	& $26.07 \pm 0.02$		&$ -3 \pm 1$	& $ -$				&	$-$ \\ 
\hline
 Diagonal			& $(26.07)$				&$ -$			& $16.5 \pm 0.2$	&$ 0.049 \pm 0.001$ \\ 
\hline
 Full fit			&$ 26.17 \pm 0.01$		&$ 7 \pm 1$		& $15.0 \pm 0.1$	&$ 0.046 \pm 0.001$ \\ 
\hline
 Previous method*	&$ 23 \pm 1$				& $-	$		& $13.9 \pm 0.7$	& $0.027 \pm 0.003$ \\
\hline
		\end{tabular}
	\end{center}
	\caption{The values of $\Gamma, n_0, \mu$ and $\lambda$ 
fitted in different ways. $^*$We also present the parameters of the effective action extracted  from the covariance matrix of volume fluctuations in our earlier work \cite{we}.  }
	\label{table2}
\end{table}

\begin{figure}[t]
\begin{center}
\includegraphics[width=0.8\textwidth]{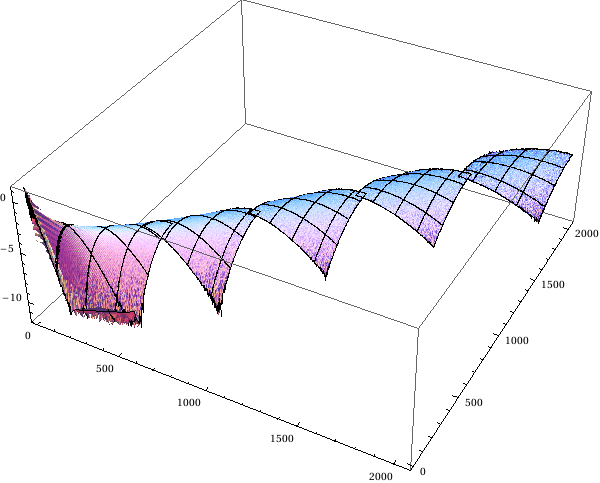}
\end{center}
\caption{Logarithm of merged and 
scaled transfer matrices measured for different $n_{vol}$.}
\label{FigMerged}
\end{figure}

Summing up, the effective action determined via the transfer matrices is 
strikingly well described by eq.\  (\ref{Eqseff}).
As in the last subsection we can merge the scaled matrices for all 
the choices of $n_{vol}$ (see Fig. \ref{FigMerged}), 
and fit expression (\ref{Eqseff}) to the aggregated data. 
The best fit 
gives $\Gamma = 26.17$, $n_0 = 7$, $\mu = 15.0$ and $\lambda = 0.046$.
We summarize the  results of fitting the parameters of $L_{eff}$ 
in different ways in Table \ref{table2}. 
As long as we are 
concentrating on the large (bulk) values of $n_t$ the various data clearly 
do not allow us to improve the expression (\ref{Eqseff}).

For comparison we also present the parameters of the effective action measured indirectly from the covariance matrix of volume fluctuations. 
This  method is based on the analysis of the effective propagator around the semi-classical solution \cite{we}. 
The parameters of the action from the previous method agree quite well with those measured directly from the transfer matrix. 
The small difference may result from slightly different parametrization of the potential term in the effective action (see footnote 4).

\subsection{Miscellaneous}

\begin{figure}[t]
\begin{center}
\includegraphics[width=0.8\textwidth]{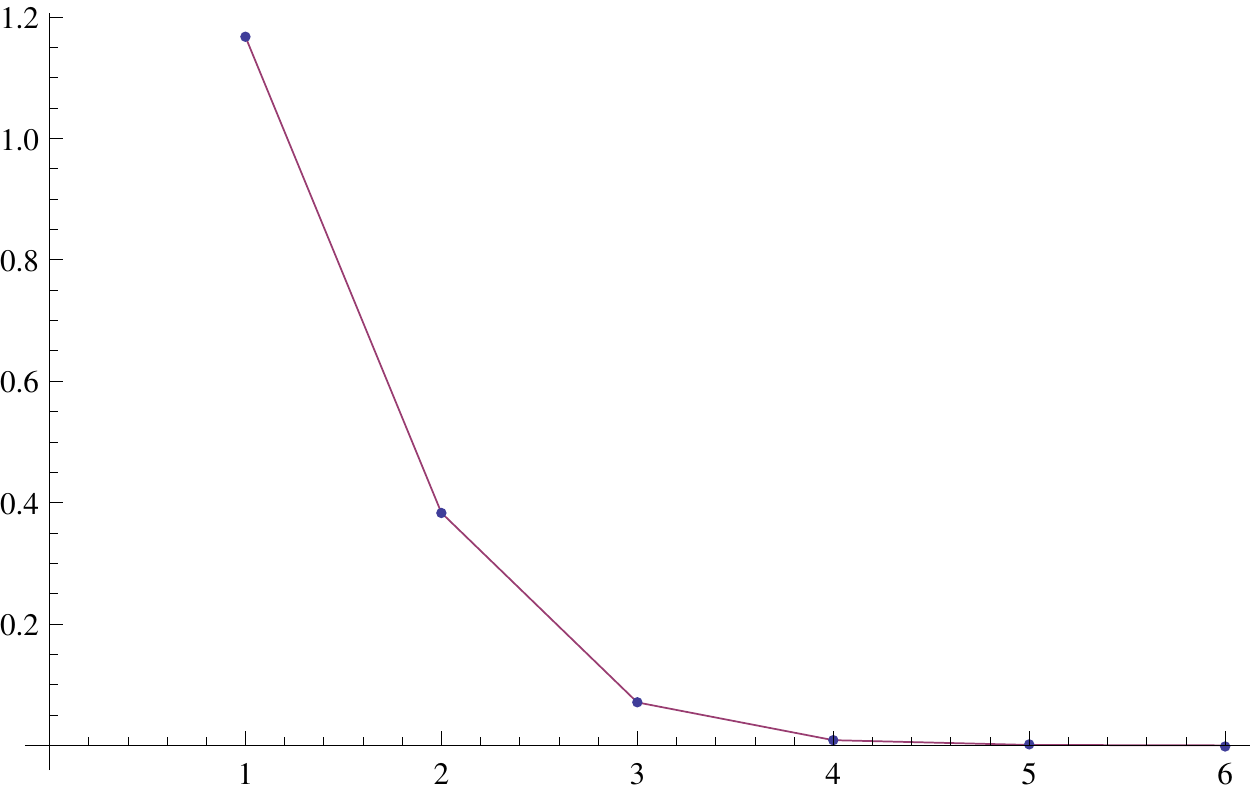}
\end{center}
\caption{The first $6$ eigenvalues of the measured transfer matrix
$\la n|M|m\ra$ calculated for $n_{vol} =  1400$ (dots)
and the similar eigenvalues for  
the fitted transfer matrix $\la n|M^{(th)}|m\ra$ (line).}
\label{FigEigenvalues}
\end{figure}

\begin{figure}[t]
\begin{center}
\includegraphics[width=0.4\textwidth]{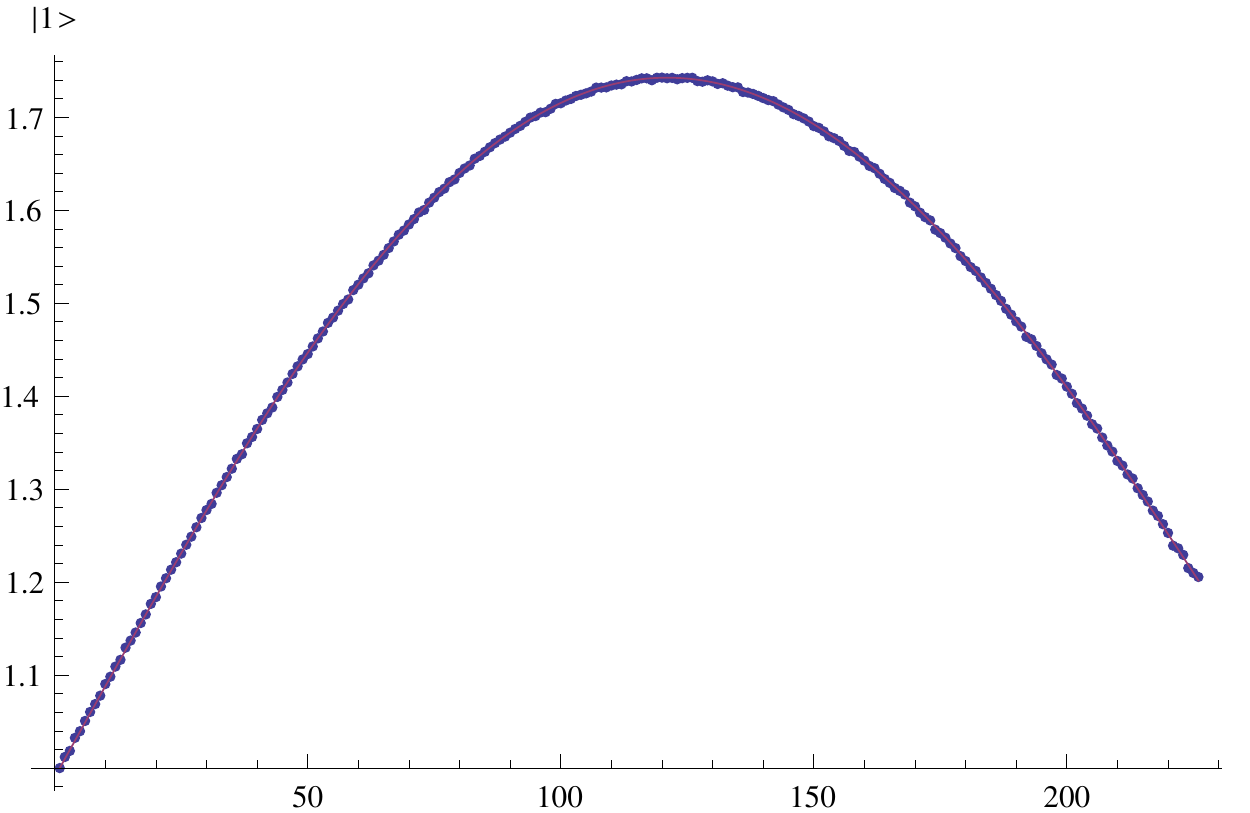}
\includegraphics[width=0.4\textwidth]{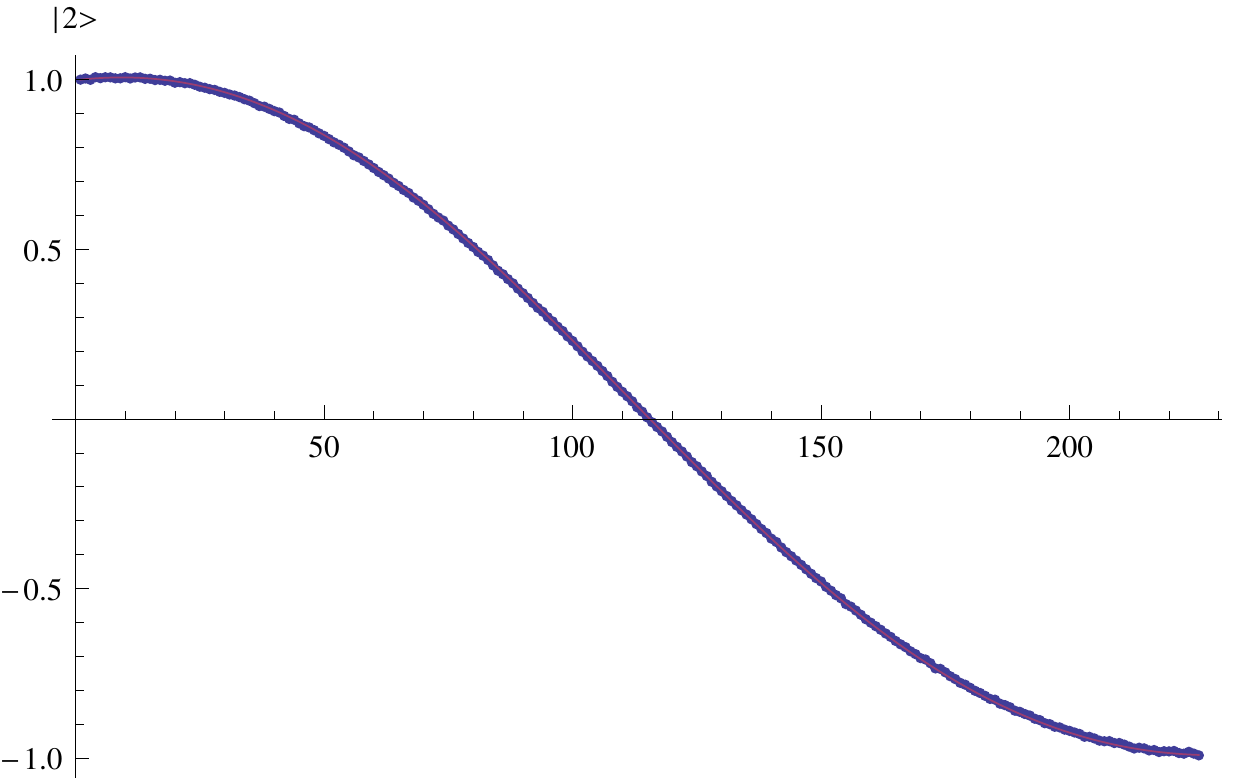}\\
\includegraphics[width=0.4\textwidth]{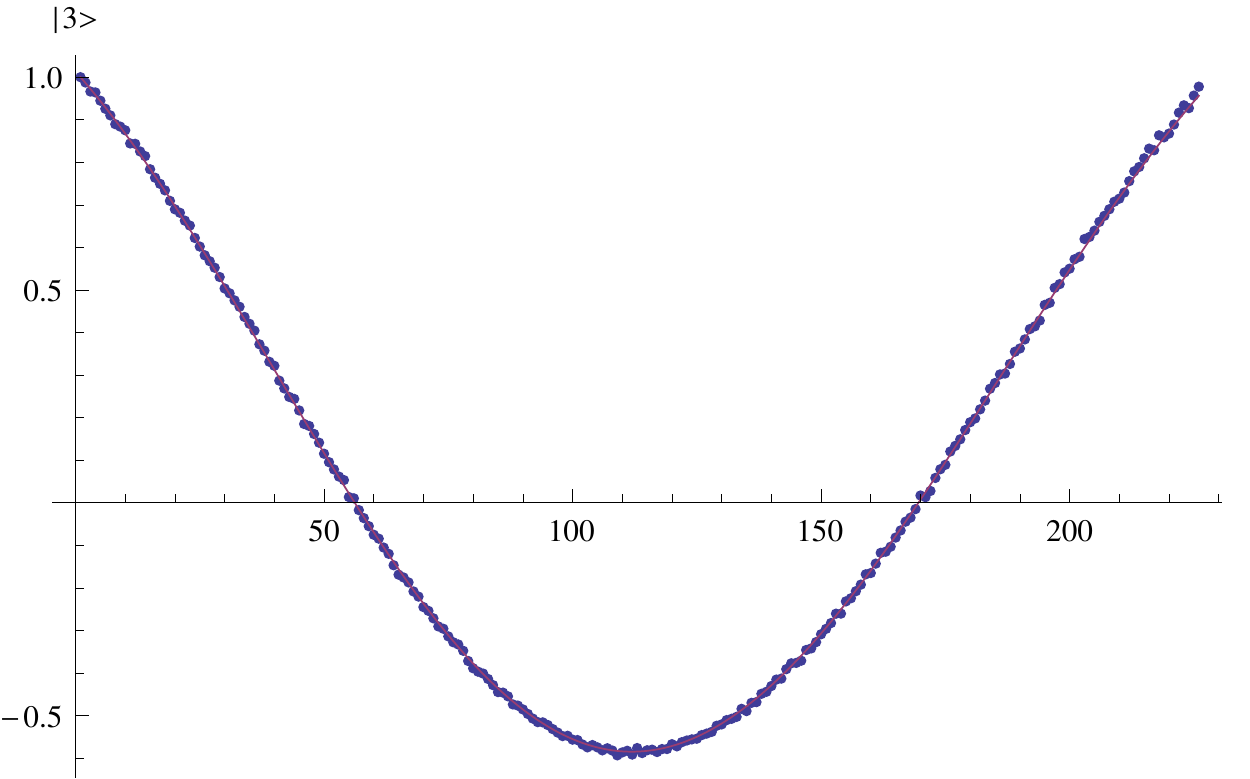}
\includegraphics[width=0.4\textwidth]{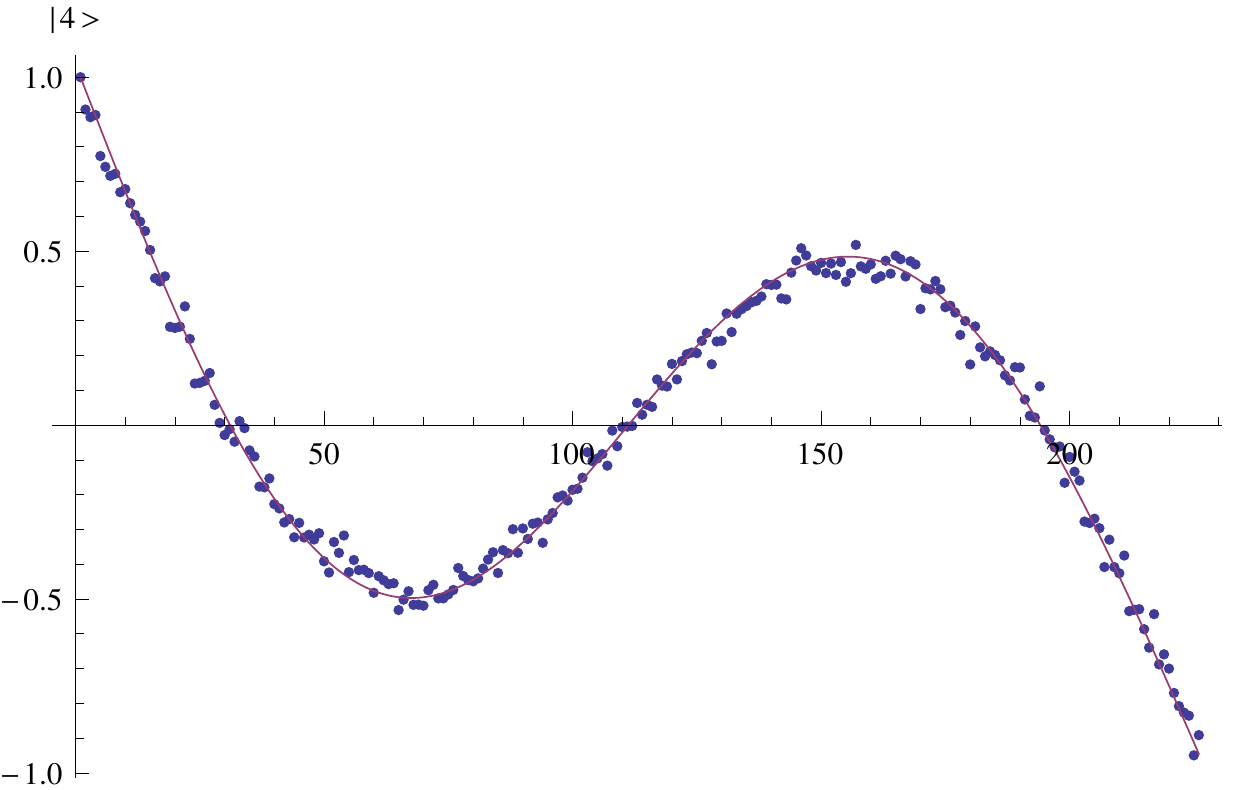}
\end{center}
\caption{The first $4$ eigenvectors of the measured 
transfer matrix $\la n|M|m\ra$ (calculated for $n_{vol} =  1400$ (dots)),
and the first 4 eigenvectors of the corresponding theoretical transfer matrix 
$\la n|M^{(th)}|m\ra$ (lines).}
\label{FigEigenvectors}
\end{figure}

The spectral decomposition of matrices presents us with 
an interesting  way to compare the measured transfer matrix 
$\la n|M|m\ra$ with the ``theoretical'' matrix $\la n|M^{(th)}|m\ra$ 
given by eq.\ (\ref{EqMth}) with parameters obtained from a best fit, 
as described above (in this case the fit from the $n_{vol} =  1400$
data). We obtain very good agreement between eigenvalues and eigenvectors 
for the two matrices. The first $6$ eigenvalues 
are presented in Fig.\ \ref{FigEigenvalues}
and the first $4$ eigenvectors are shown in Fig.\ \ref{FigEigenvectors}.

\begin{figure}[t]
\begin{center}
\includegraphics[width=0.45\textwidth]{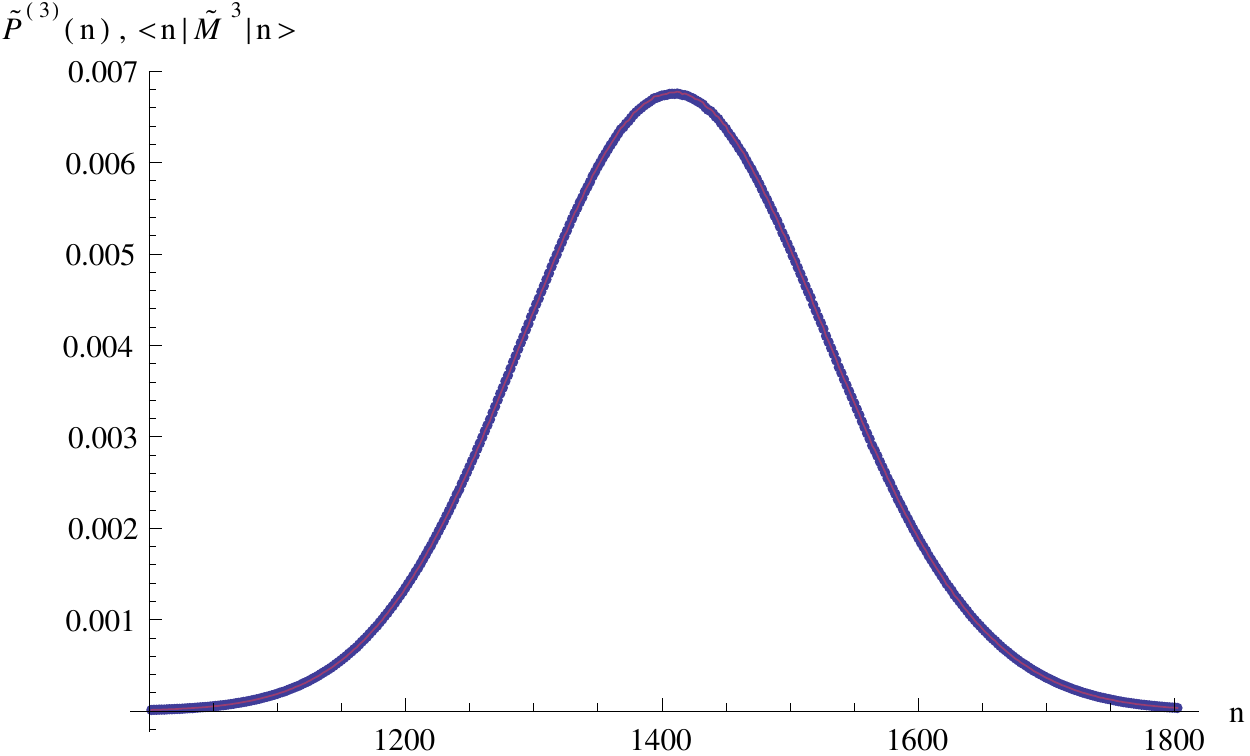}
\includegraphics[width=0.45\textwidth]{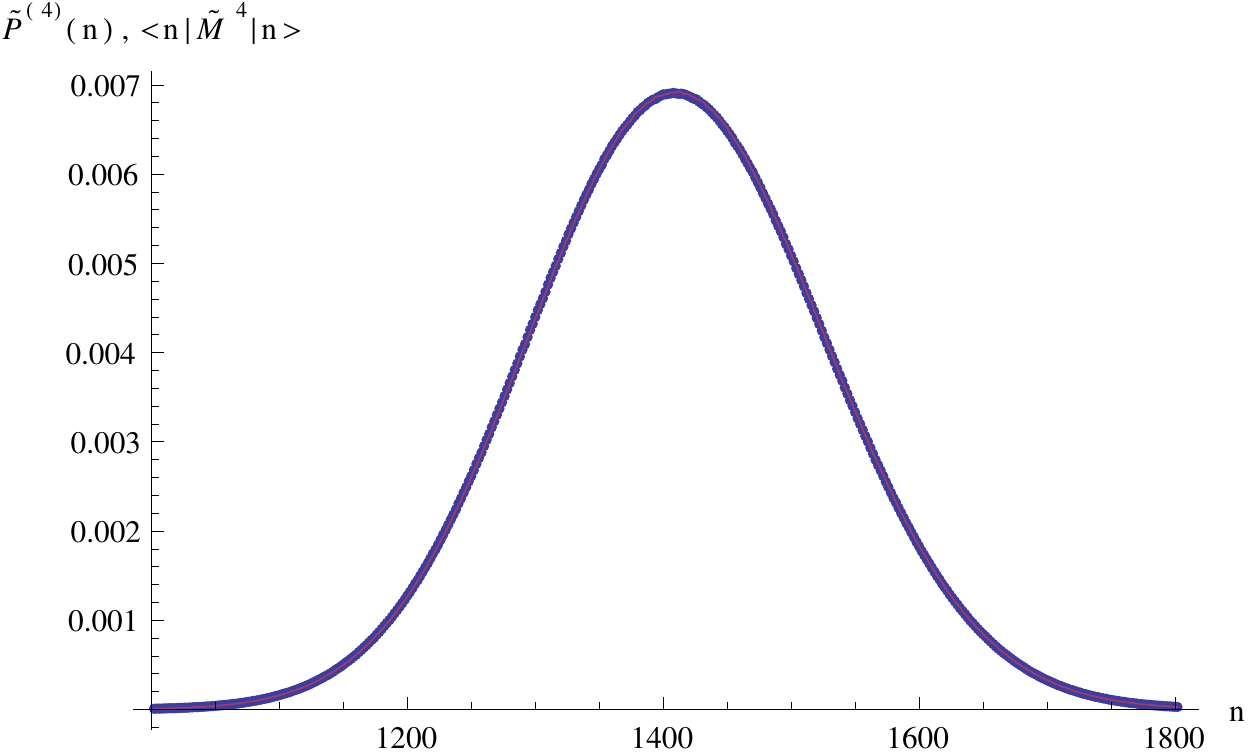}
\end{center}
\caption{
The probability distribution $\tilde{P}^{t_{tot}}(n)$ (dots) 
and $\frac{\langle n | \tilde{M}^{t_{tot}} | 
n \rangle}{\tr \tilde{M}^{t_{tot}}}$ (line)
for $t_{tot}=3$ (left) and $t_{tot} = 4$ (right).}
\label{FigHist34}
\end{figure}

As a final check of the consistency of the effective transfer matrix
with data we measure the probability distribution of a spatial slice volume 
$\tilde{P}^{(t_{tot})} (n)$ in terms of the transfer matrix $\tM$ given by
\begin{equation}
	\tilde{P}^{t_{tot}} (n) = 
\frac{\langle n | \tilde{M}^{t_{tot}} | n \rangle}{\tr \tilde M^{t_{tot}}}.
	\label{EqPT1Check}
\end{equation}
We use $\tilde{M}$ instead $M$, because 
the power $M^{t_{tot}}$ involves summation over all possible volumes $n$
which are not accessible until we suppress them with 
the term $e^{-\epsilon (n-n_{vol})^2}$ 
(which is present in $\tilde{M}$).
We compare the theoretical expectation (\ref{EqPT1Check}) 
with the  measured $\tilde{P}^{t_{tot}} (n)$
for $t_{tot} = 3$ and $t_{tot} = 4$.
The comparison is shown on Fig. \ref{FigHist34} 
and the error is of order $0.02\%$.

\section{The transfer matrix for small three-volumes}

We have seen that the effective transfer matrix is 
very well described by the simplest effective Lagrangian \rf{Eqseff}.
However, it is natural to expect that the corresponding effective
action is only a first approximation. Appealing to 
isotropy and homogeneity one would expect that 
a potential $R^k(t)$ term, where $R(t)$ refers to the scalar
curvature of the three-dimensional space at time $t$,
translates into a term $ \left( \frac{n_t + n_{t+1}}{2} \right)^{1-2k/3}$. The leading term
$R(t)$ is already part of the effective action where it 
appears as the term $\left( \frac{n_t + n_{t+1}}{2} \right)^{1/3}$. For the data coming from
large $n_t$ we have seen that the difference between 
our measured $M$ and $M^{(th)}$ coming from the effective 
action with only the $\left( \frac{n_t + n_{t+1}}{2} \right)^{1/3}$ term is already at the noise 
level.  From these large $n_t$ data we have no chance with the present 
statistics to study higher $k$ corrections. Thus we now turn 
to the small $n_t$ region. The difficulty of analyzing 
the small $n_t$ region is that we might encounter 
discretization effects as is apparent from Fig. \ref{Fig02}.
The study of this region started in \cite{we}, but the 
present approach offers the great advantage that we can 
perform high statistics study of small systems, while in the 
earlier studies the interesting region was a small  part of 
a larger system and thus the statistics becomes less good.

We want to measure the effective transfer matrix 
by measuring $P^{t_{tot}}(n_i,n_j)$ for $t_{tot}=3,4$
as we have already done, but now for small $n_t$, i.e.\
for much smaller systems. We also want to check as well 
as possible that the concept of an ``effective'' transfer matrix
actually works. While it seemed to work well for large $n_t$, it 
is not obvious that this will remain true for small $n_t$: physics
might be different and discretization effects might also spoil such
a picture. Our systems will be so small that we do not have 
to introduce the auxiliary transfer matrix $\tM$ and the 
volume fixing parameter $n_{vol}$. Thus we only use 
the Regge Einstein-Hilbert action $S_R$, eq.\ \rf{Sdisc}, to generate
the probability distributions $P^{(3)}(n_1,n_2)$ and $P^{(4)}(n_1,n_3)$ 
and construct $\la n|M| m \ra$ from eq.\ \rf{ja13d}.

\subsection{Eigenvectors analysis.}

Given the transfer matrix $M$ we can perform a spectral decomposition
in terms of eigenvalues $\lam_i$ and (orthonormal) eigenvectors 
$\ket{\alpha_i}$:
\begin{equation}
M = \sum_{i} \lambda_i \ket{\alpha_i}\bra{\alpha_i}.
\label{Meigenvectors}
\end{equation}
Since the measured $M$ is only determined up to a normalization,
we will assume $\lam_1=1$ and  $| \lambda_1 | \geq | \lambda_2 |\geq ...$.

If the gaps are significant between the first,   second and  third 
eigenvalues , it is clear that the large $t_{tot}$ limit of 
$\cZ(t_{tot})$ and the large $\Del t$ 
limit of the simplest correlation functions 
$P^{t_{tot}}(n_t,n_{t+\Del t})$ will
be completely dominated by the first two eigenstates: 
Thus, in the limit where $t_{tot} \gg \Del t \gg 1$  we have
(recalling that $\lam_1$ is normalized to 1)
\begin{equation}
\cZ(t_{tot}) = \sum_i \left(\lambda_i\right)^{t_{tot}} \approx 1  
\label{Zeigen}  
\end{equation}
\begin{equation}
P^{t_{tot}}(n) =  \frac{1}{Z} \sum_i \lambda_i^{t_{tot}}  
\braket{n|\alpha_i}^2 \approx \braket{n|\alpha_1}^2 
\label{p1eigen}
\end{equation}
\bea
P^{t_{tot}}(n_t,m_{t+\Del t}) &=& 
\frac{1}{Z} \Big( \sum_i \left(\lambda_i\right)^{\Delta t} 
\braket{n|\alpha_i} \braket{\alpha_i|m}\Big)       
\Big( \sum_i \left(\lambda_i\right)^{t_{tot}-\Delta t} 
\braket{n|\alpha_i} \braket{\alpha_i|m}\Big)
\nonumber \\
 &\approx&     \braket{n|\alpha_1}^2 \braket{\alpha_1|m}^2
 +    \lambda_2^{\Delta t} \braket{n|\alpha_2} 
\braket{\alpha_2|m} \braket{n|\alpha_1} \braket{\alpha_1|m} 
\label{p2eigen}\nonumber
\eea
The average can be written as:
\begin{equation}\label{average}
\la n\ra_{t_{tot}} \approx \sum_n n \braket{n|\alpha_1}^2
\end{equation}
For the correlator we have:
\begin{equation}\label{correlator}
\la n_t m_{t+\Del t}\ra -\la n_t\ra \la m_{t+\Del t}\ra \approx 
\lambda_2^{\Delta t} \sum_{n,m} n m \braket{n|\alpha_2} 
\braket{\alpha_2|m} \braket{n|\alpha_1} \braket{\alpha_1|m}. 
\end{equation} 
and the long distance behavior is an exponential fall off 
$e^{-\mu \Del t}$, $\mu = -\log \lam_2/\lam_1$ (where we have reintroduced
$\lam_1$ for clarity).

\begin{figure}[!ht]
 \centering
  \scalebox{0.8}{\includegraphics{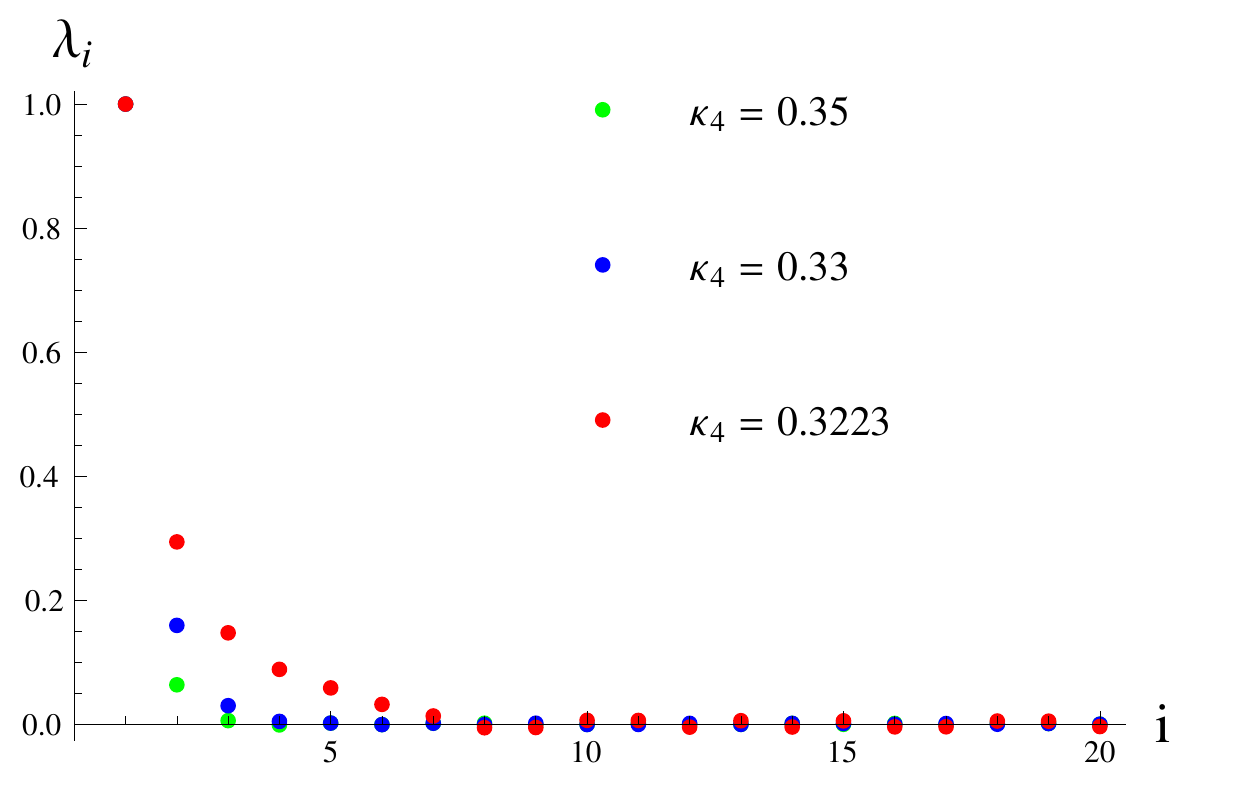}}
\caption{Eigenvalues of the transfer matrix $M$ for different values of 
$\kappa_4$ approaching $\kappa_4^{crit}\approx 0.3222$. 
Note that according to our normalization 
the biggest eigenvalue is set to one.} 
\label{fig1}
\end{figure}
How well are these approximate relations satisfied? First we observe
that there is indeed a clear gap between the first eigenvalues, as
illustrated in Fig.\ \ref{fig1}. The closer $\kp_4$ is fine-tuned
to $\kp_4^{crit}$ the smaller the gap, but even very close to the
critical value we observe a clear gap. That we have gaps 
even at $\kp_4^{crit}$ just illustrates that we indeed consider  small systems.

\begin{figure}[ht]
    \centering
  \scalebox{0.9}{\includegraphics{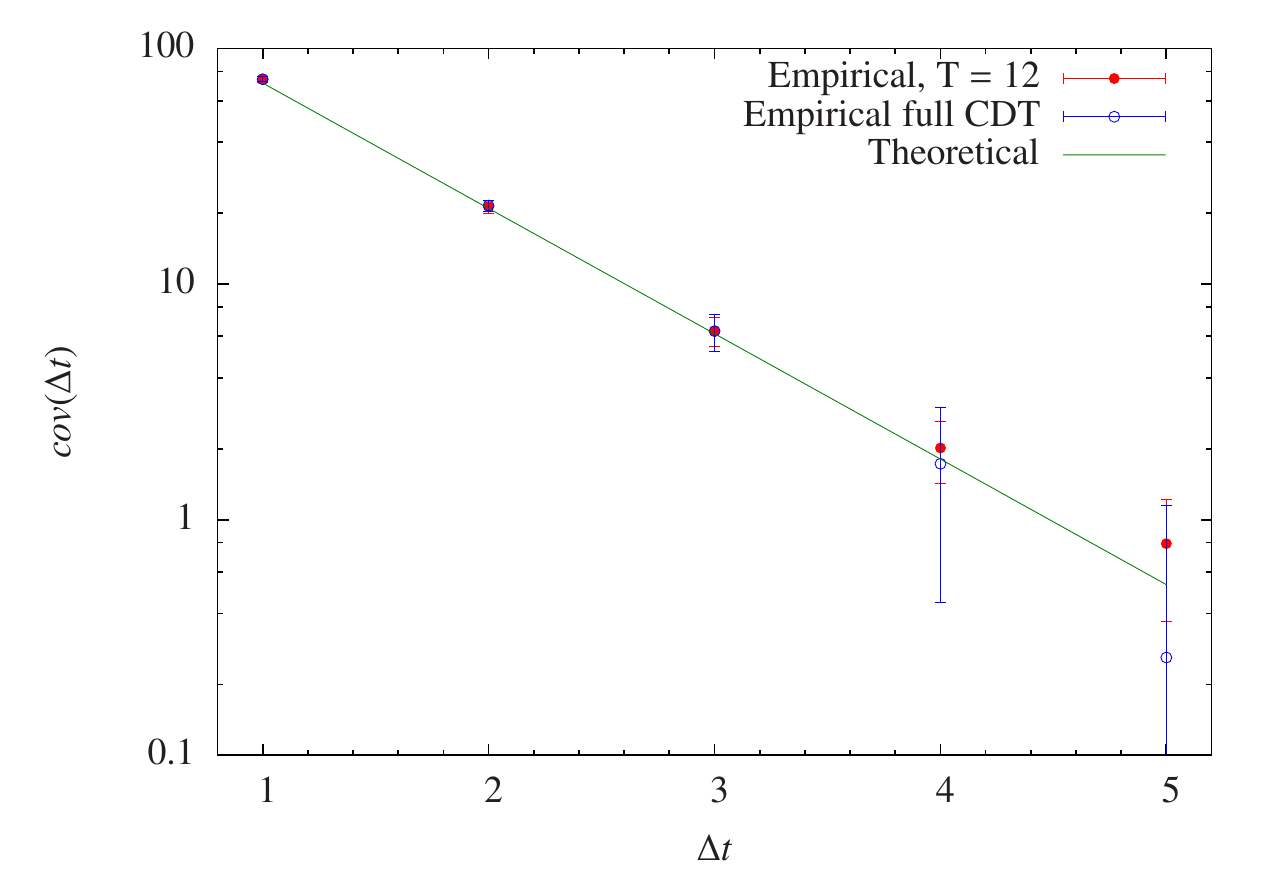}} 
\caption{The ``theoretical'' correlator \rf{correlator} of spatial volumes 
between time-slices separated by $\Delta t$ and  
calculated including the first two eigenvectors of 
the transfer matrix $M$ calculated for $\kappa_4 = 0.3223$ (green line). The correlator decays exponentially with $\Delta t$ (log scale). There is  a very good agreement with the correlator measured directly in simulations for T=12 
(red) and in ''full CDT'' stalk range (blue). The bars indicate measurement errors. }
    \label{cov}
\end{figure}

Next we ask how large $t_{tot}$ has to be in order that 
the approximations made in eq.\ \rf{p1eigen}-\rf{correlator} are 
valid. That can of course be read off from the eigenvalues and 
already for $t_{tot} =4$ the approximation is very good for 
$\la n \ra_{t_{tot}}$. For the correlator one has of course to consider
larger $t_{tot}$. In Fig.\ \ref{cov} we have shown for $t_{tot}=12$
the expected exponential decay with exponent $\log (\lam_2/\lam_1)$ 
compared to the actually measured correlator. The agreement is 
very good even for small $\Del t$ where it is not obvious that 
ignoring the eigenvectors with eigenvalues smaller than $\lam_2$
is a valid approximation.

\subsection{The ``Full-CDT'' approximation.}

\begin{figure}[!ht]
    \centering
  \scalebox{0.9}{\includegraphics{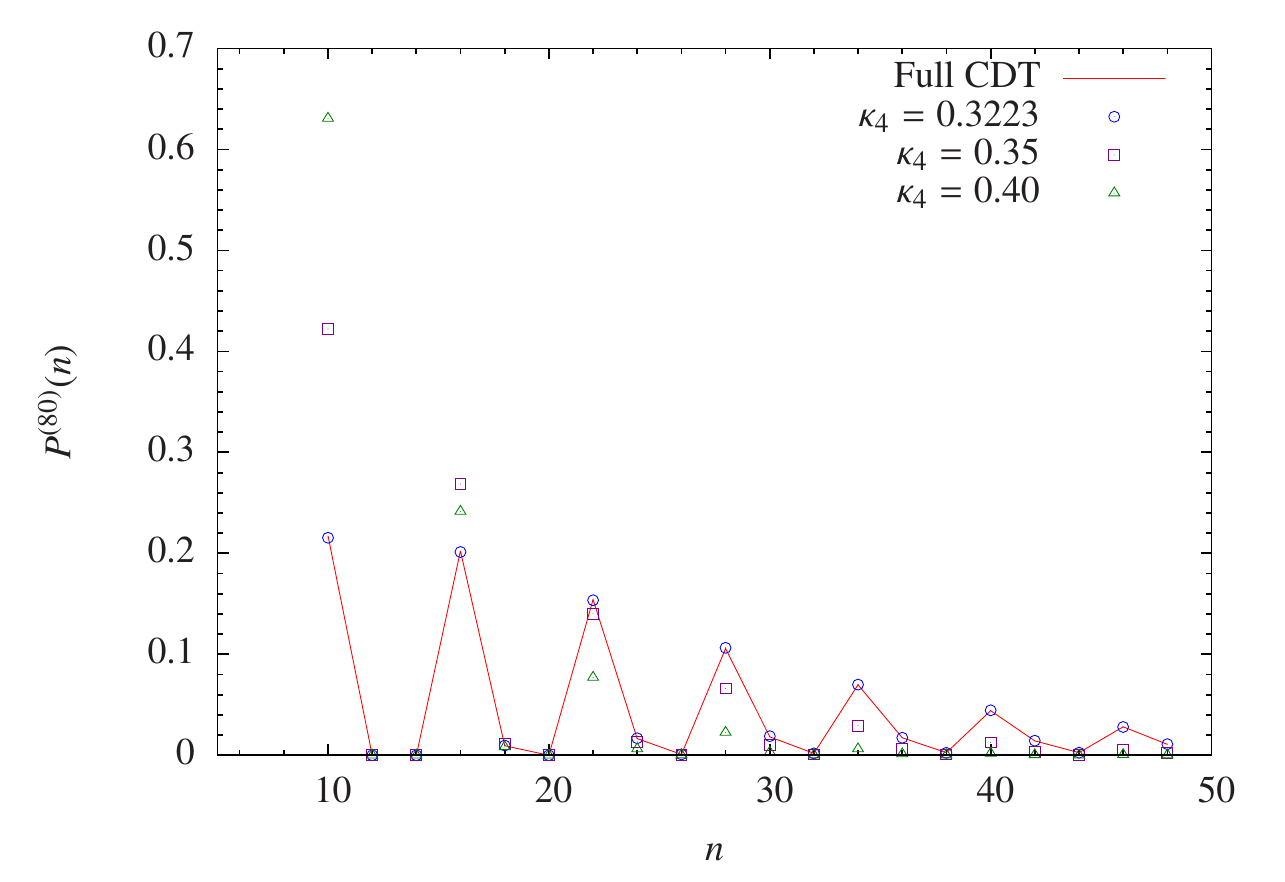}}
\caption{Theoretical 1-point probability distributions calculated 
with the 'largest' eigenvector of the transfer matrix $M$ approach 
empirical probability measured in ''full CDT'' (red line) as $\kappa_4$ 
tends to critical value. Theoretical distributions were computed 
for $\kappa_4$ = $0.40$, $0.35$ and $0.3223$. For $\kappa_4 = 0.3223$ (blue circles) the agreement is very good.} 
    \label{fig6}
\end{figure}

We have also compared our approximate large $t_{tot}$ probability 
distributions $P(n)$ (eq. \ref{p1eigen}) with the data taken from the stalk 
range of \rm{full CDT} simulations (including the blob and the stalk range).
The distributions  
approach very well ``full CDT'' measurements as $\kappa_4$ tends to 
critical value (Fig. \ref{fig6}). The validity of the transfer matrix model  is further confirmed by the behaviour of the correlator $\la n_t m_{t+\Del t}\ra -\la n_t\ra \la m_{t+\Del t}\ra $ measured directly in the stalk range of ''full CDT''. As illustrated in Fig.\ \ref{cov} the measured correlator falls off as  $e^{-\mu \Delta t}$. The parameter $\mu$ is well explained by the ratio of the first two eigenvalues of the transfer matrix $M$ calculated for $\kappa_4$ closest to the critical value. 

\section{The effective action for small three-volumes}

In principle the matrix $M$ presented above would allow us 
to determine an effective Lagrangian for small $n_t$, i.e.\ even in
the stalk range of the CDT configurations, precisely as we did 
for large $n_t$:
\begin{equation}
S_{eff} = \sum_{t} L_{eff}(n_t,n_{t+1}),\quad 
\braket{n | M | m} = {\cal N} e^{-L_{eff}(n,m)}.
\label{seff}
\end{equation}
However, we are confronted with  the existence of three families of states, 
as is apparent in Fig.\ \ref{Fig02}. We can however define
a {\it reduced} matrix $\hat{M}$ 
performing a summation over the three families, i.e.
\begin{equation}
\hat{M} = U M U^T
\end{equation} 
where the rectangular matrix $U$ has a form:
$$
 U = \left\{\begin{array}{cccccccccc} 
1 & 1 & 1 & 0 & 0 & 0 & 0 & 0 & 0 &  \cdots \\
 0 & 0 & 0 & 1 & 1 & 1 & 0 & 0 & 0 & \cdots \\
0 & 0 & 0 & 0 & 0 & 0 & 1 & 1 & 1 & \cdots\\
\cdots \end{array}\right\}
$$
The elements of the matrix $\hat{M}$ behave much more smoothly 
and can be analyzed, using the effective action idea. 
We normalize $\hat{M}$ by choosing its largest eigenvalue to be one.

Let us assume that  $L_{eff}$ has the form:
\begin{equation}
L_{eff}(n,m)=\frac{(n-m)^2}{k(n+m)}+v(n+m)
\label{seffform}
\end{equation}
where the functions $k(\cdot)$ and $v(\cdot)$ are to be determined.

\begin{figure}[!ht]
\centering
\scalebox{0.7}{\includegraphics{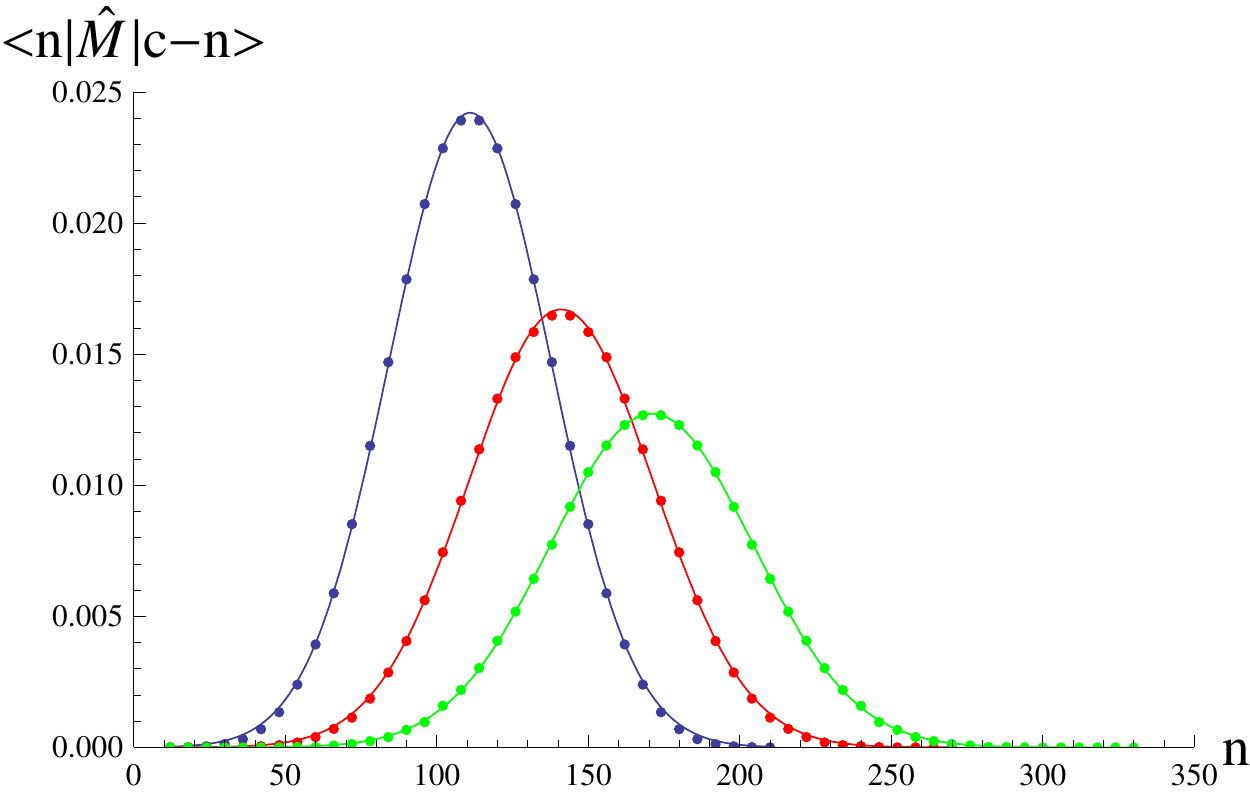}}
\caption{$\braket{n|\hat M|c-n}$ plotted as a function of $n$ for $c=222$ (blue), $c=282$ (red), $c=342$ (green). Gaussian fits (\ref{gauss}) are presented as solid lines. }
\label{Mgauss}
\end{figure}
\begin{figure}[!ht]
\centering
\scalebox{0.7}{\includegraphics{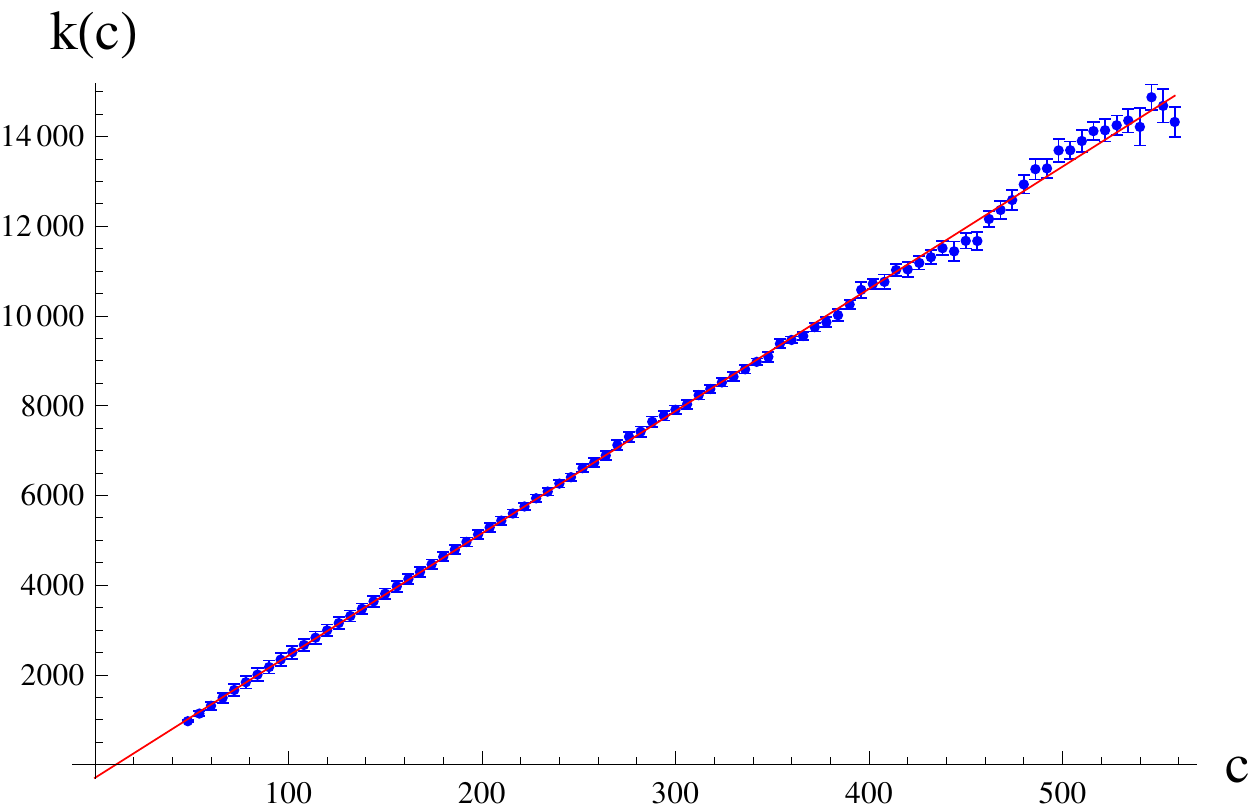}}
\caption{Kinetic coefficients $k(c)$ (blue) and a linear fit (red).}
\label{kinetic}
\end{figure}

We follow the same strategy as when $n_t$ was large: first 
we analyze matrix elements  for constant $n+m=c$ in order 
to keep the potential term $v(n+m)$ constant. 
One observes a Gaussian dependence on $n$ (Fig. \ref{Mgauss}):
\begin{equation}
\braket{n | \hat M | m} = \braket{n | \hat M | c - n}=  
{\cal N} \exp\left[-\frac{(2n-c)^2}{k(c)}-v(c)\right]
\label{gauss}
\end{equation} 

Fitting (\ref{gauss}) for different $c$'s one can 
easily check that the kinetic coefficient $k(c)$ 
is linear (Fig. \ref{kinetic}). Therefore one can write:
\begin{equation}
L_{eff}(n,m) = \frac{1}{\Gamma}\left[ \frac{(n-m)^2}{n + m - 2 n_0}+
\widetilde v(n+m)\right]
\label{kin}
\end{equation}
where: $\widetilde v() = \Gamma v()$. The best fit of $\Gamma$ 
and $n_0$  is presented in Table \ref{table3}.

In the above we recognize  the familiar 
kinetic term present in the effective action for the blob 
range. Let us also test the assumption that the potential 
part is similar by analyzing the diagonal elements of $\hat M$
where the kinetic term is zero. From (\ref{seff}) and (\ref{kin}) one obtains
\begin{equation}
 \log  \braket{n |\hat M | n} = - \frac{1}{\Gamma} \widetilde v(2n) +  
\log {\cal N}
\end{equation}

\begin{figure}[!ht]
\centering
\scalebox{0.7}{\includegraphics{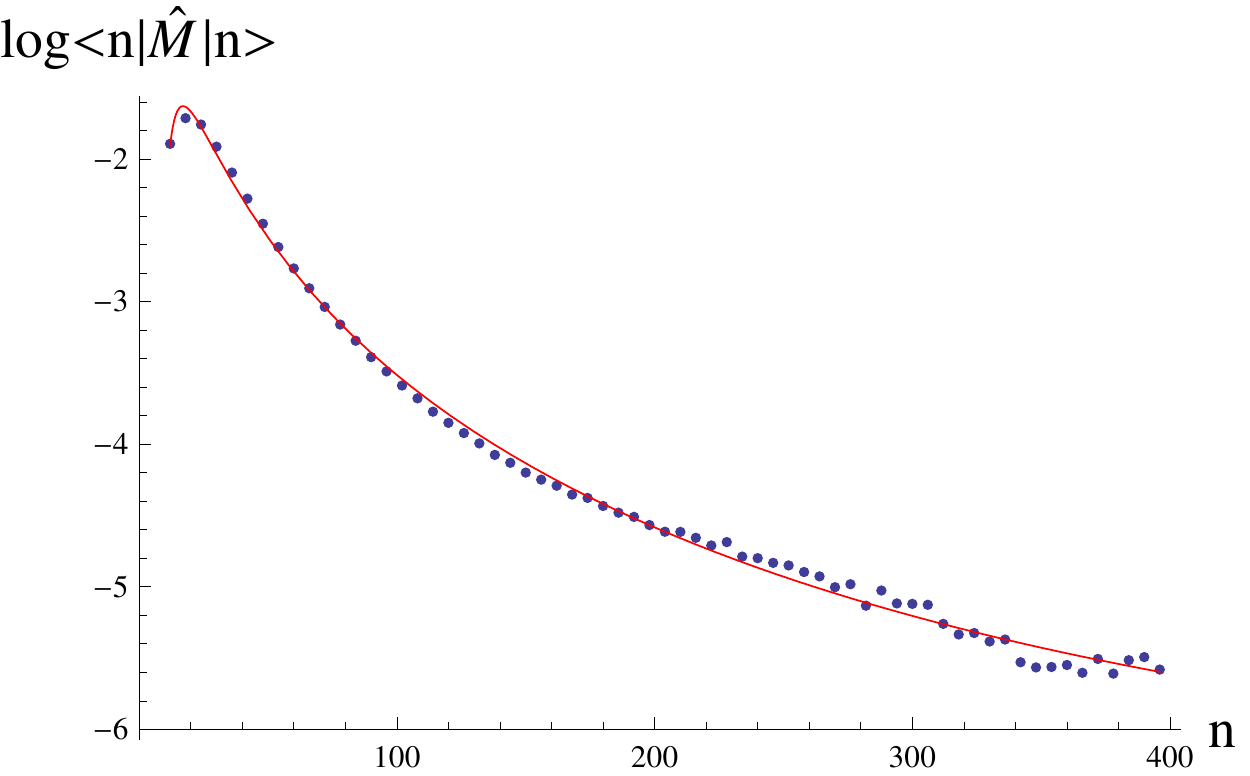}}
\caption{Logarithm of the diagonal $\braket{n|\hat M |n}$ (blue dots) 
compared with the fit of the potential term (\ref{pot}) (red line).}
\label{diagonal}
\end{figure}

Let us assume 
\begin{equation}
\widetilde v(2n)= 
 - \lambda  n + \mu n^{1/3} + \delta n ^{-\rho} ,
\label{pot}
\end{equation}
inspired by the bulk expressions already derived and where 
we have included a new term $ \delta n ^{-\rho}$, inspired by earlier
remarks about effective powers of the three-dimensional curvature $R(t)$. 
We present the fit of the logarithm of the diagonal elements  
$\braket{n |\hat M | n}$  in Fig. \ref{diagonal} 
and the parameters of the fit in Table \ref{table3}.

\vspace{12pt}

\begin{table}[ht]
\begin{center}
\begin{tabular} {|c|c|c|}
\hline
{Parameter} & {Stalk}	& {Blob}  \\ \hline
\hline
$\Gamma$&$	27.2 \pm 0.1 $&$	25.7 - 26.2$ \\ \hline
$n_0$&$	5 \pm 1$&	$-3 - +7$  \\ \hline
$\mu$&$	34 \pm 2 $&$	13 - 30$ \\ \hline
$\lambda$&$	0.12 \pm 0.02$&$	0.04 - 0.07$ \\ \hline
$\delta$&$	(4 \pm 7)\times 10^4$&	$-$ \\ \hline
$\rho$&$	3 \pm 1$&$-$ \\ \hline
\end{tabular}
\end{center}
\caption{ Fitted parameters of the effective action for 
the stalk (\ref{Seffform}). For comparison we also present 
estimates of parameters of the effective action for the blob 
calculated from the large $n_t$ simulations 
(see Table \ref{table1} and \ref{table2}).}
\label{table3}
\end{table}


It is quite surprising that the same effective action is still 
present in the stalk, despite the volume behavior seems, at the first sight, 
quite different from that in the blob range. It is even more surprising 
that the parameters of the fit agree quite well with the effective 
Newton constant $\Gamma$ measured in the blob range. 
The parameter $\mu$ is slightly bigger but of the same order of magnitude 
as the potential coefficient from the blob range. 
Only $\lambda$ which is related to the size of the dynamically 
created universe is quite different, but that should be no surprise since 
$\lambda$ semiclassically is related to the size of the universe.
Finally the value of $\rho$ is difficult to explain from the 
point of view of higher powers of $R(t)$ which should 
give $\rho= 2k/3-1$, $k=2,\ldots$, but also it should be mentioned that 
it is not very well determined from the fits.

Summing up:
\begin{equation}
S_{eff}^{stalk}\equ \sum_t \frac{1}{\Gamma} \left[  
\frac{(n_t-n_{t+1})^2}{n_t \plu n_{t+1} \mi 2 n_0} \plu 
\mu \left(\frac{n_t\plu n_{t+1}}{2}\right)^{1/3}\! \mi 
\lambda \frac{n_t\plu n_{t+1}}{2} \plu
\delta \left(\frac{n_t\plu n_{t+1}}{2}\right)^{-\rho}\right]. 
\label{Seffform}
\end{equation}

\begin{figure}[!ht]
\centering
\scalebox{0.7}{\includegraphics{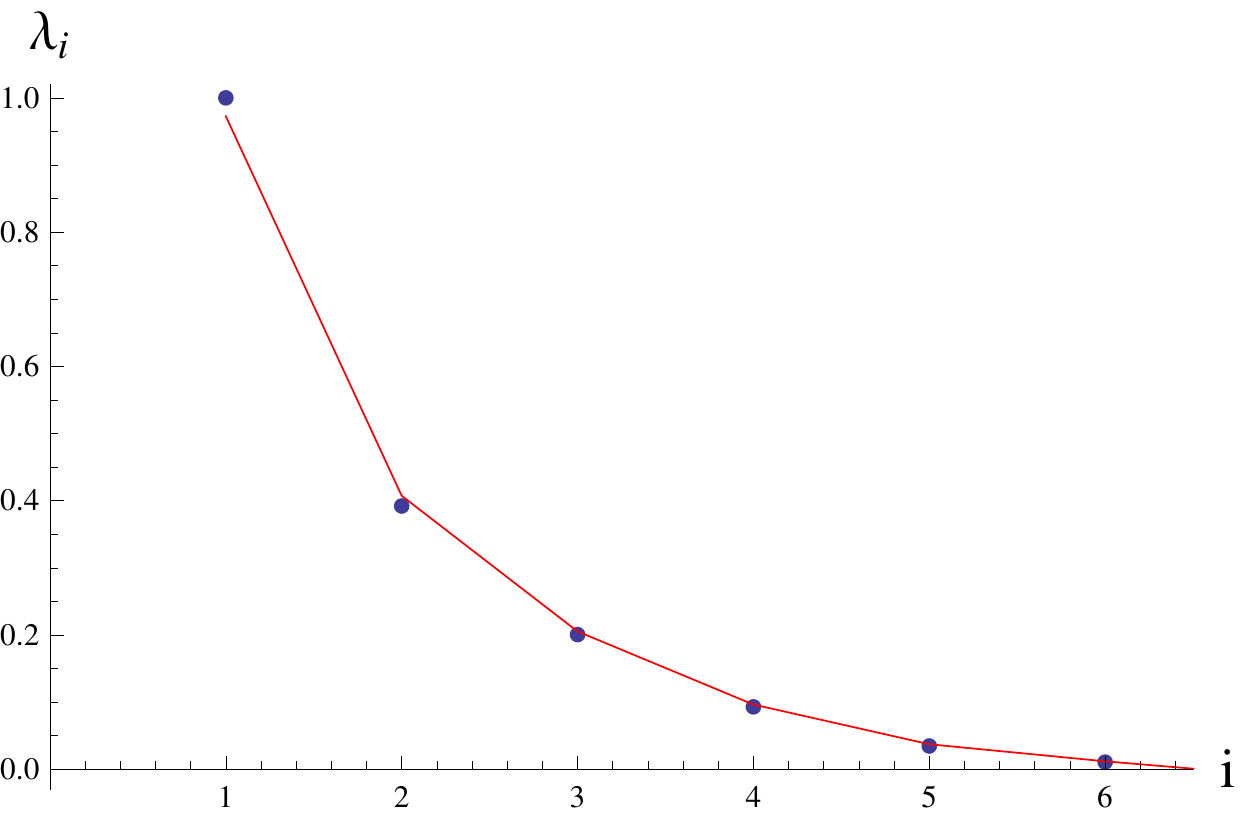}}
\caption{First six eigenvalues of the measured $\hat M$ (blue dots) 
and the theoretical matrix ${\hat M}^{(th)}$ (red lines). 
${\hat M}^{(th)}$ was calculated using the effective action 
for the stalk range (\ref{Seffform}). }
\label{eigen}
\end{figure}

\begin{figure}[!ht]
\centering
\scalebox{0.3}{\includegraphics{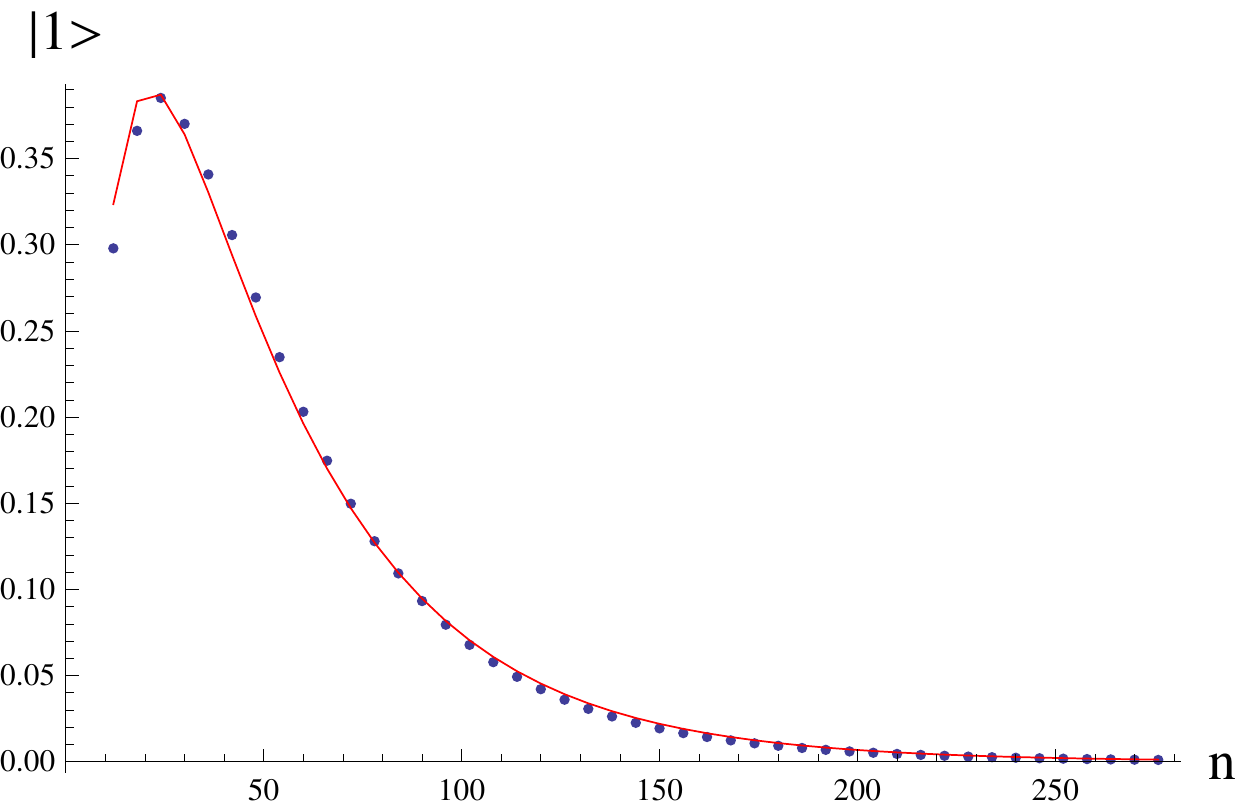}}
\scalebox{0.3}{\includegraphics{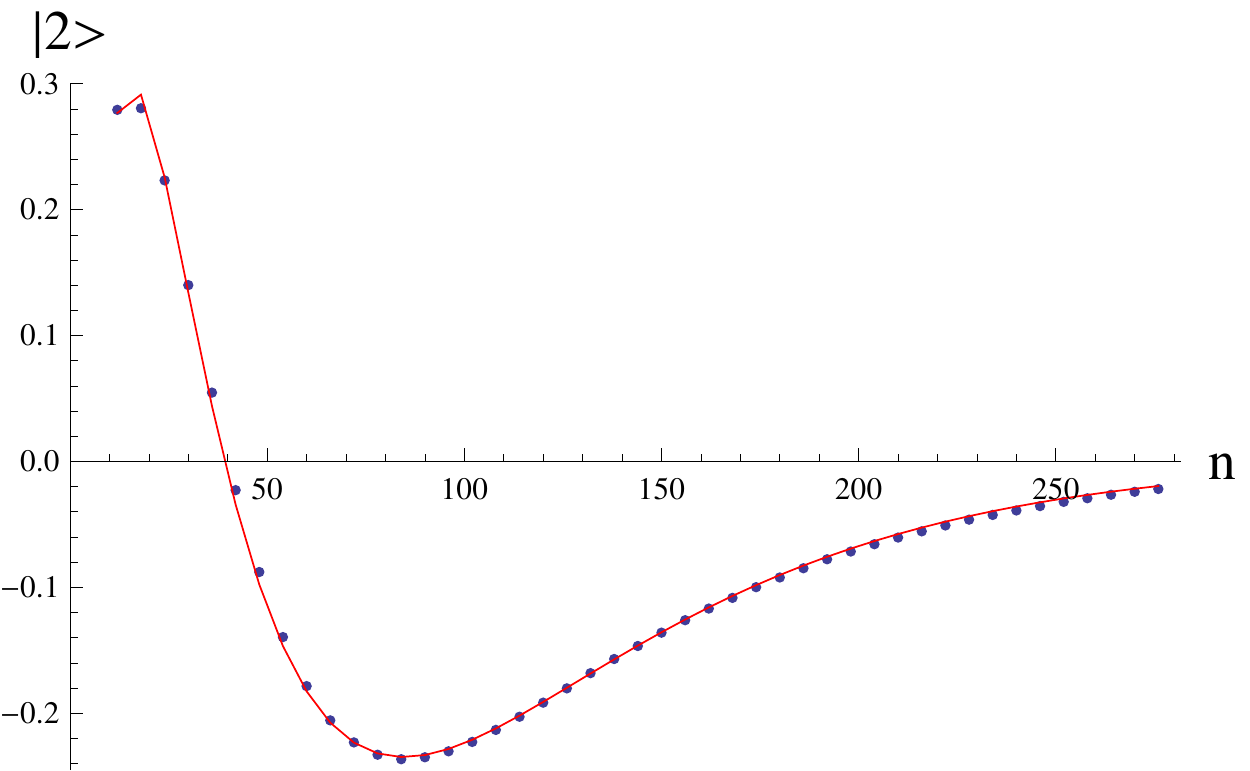}}
\scalebox{0.3}{\includegraphics{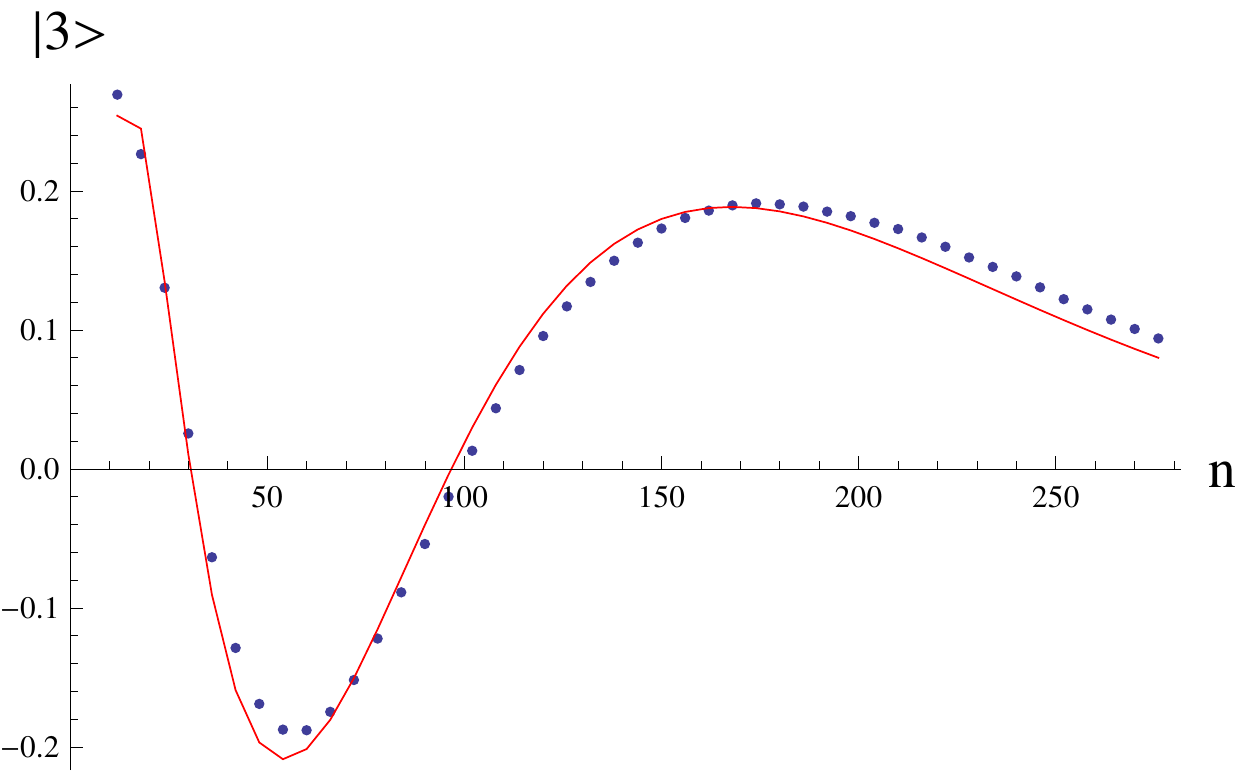}}
\scalebox{0.3}{\includegraphics{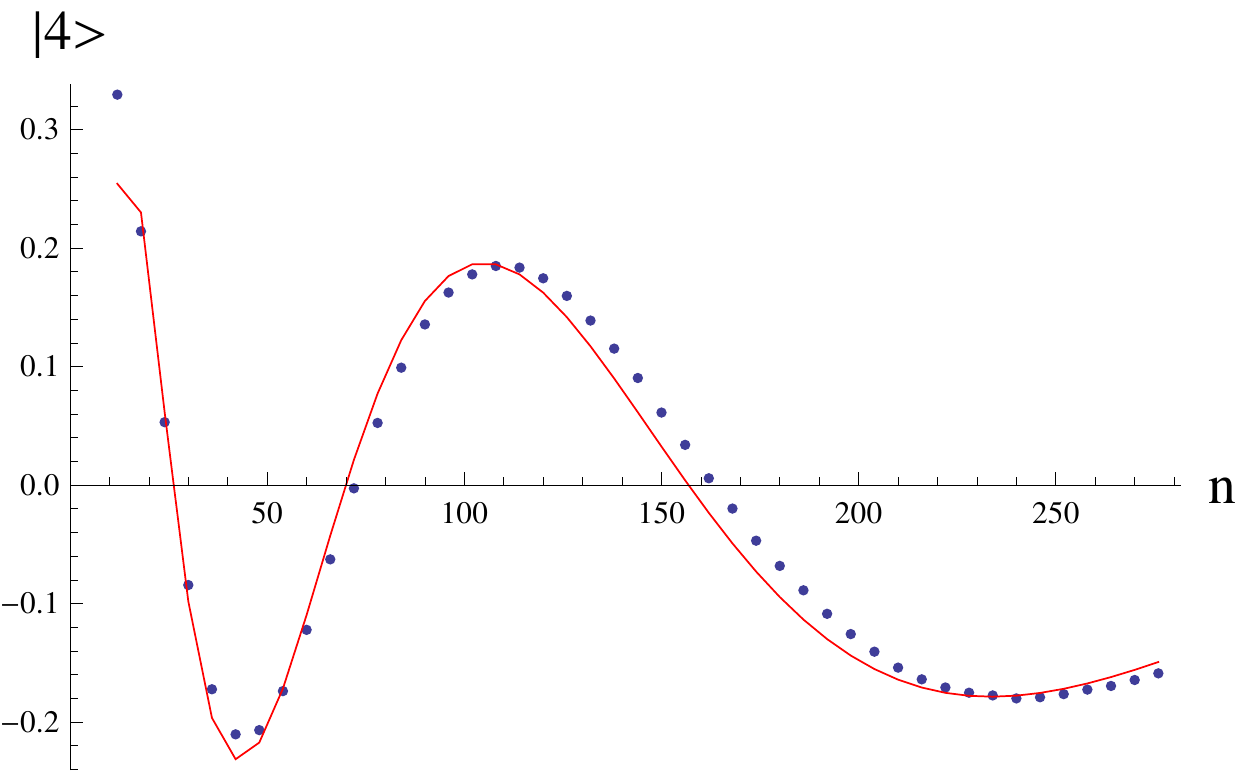}}
\scalebox{0.3}{\includegraphics{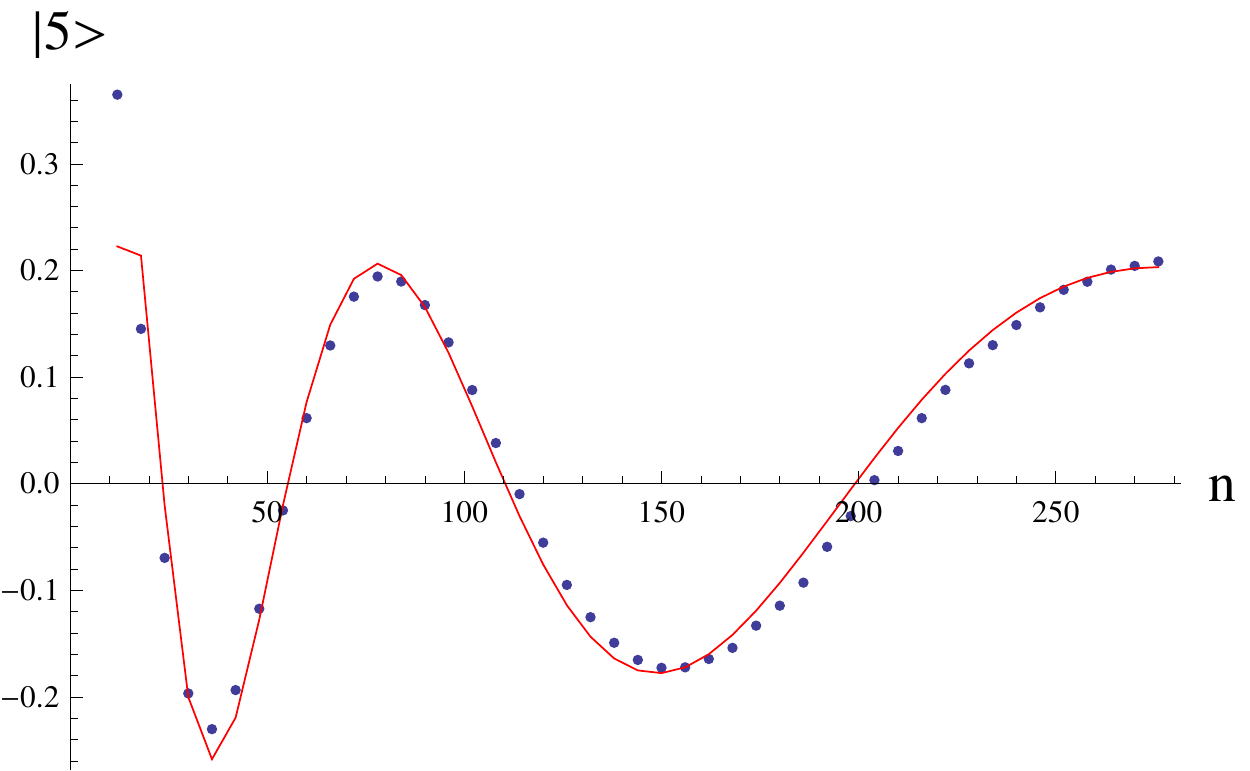}}
\scalebox{0.3}{\includegraphics{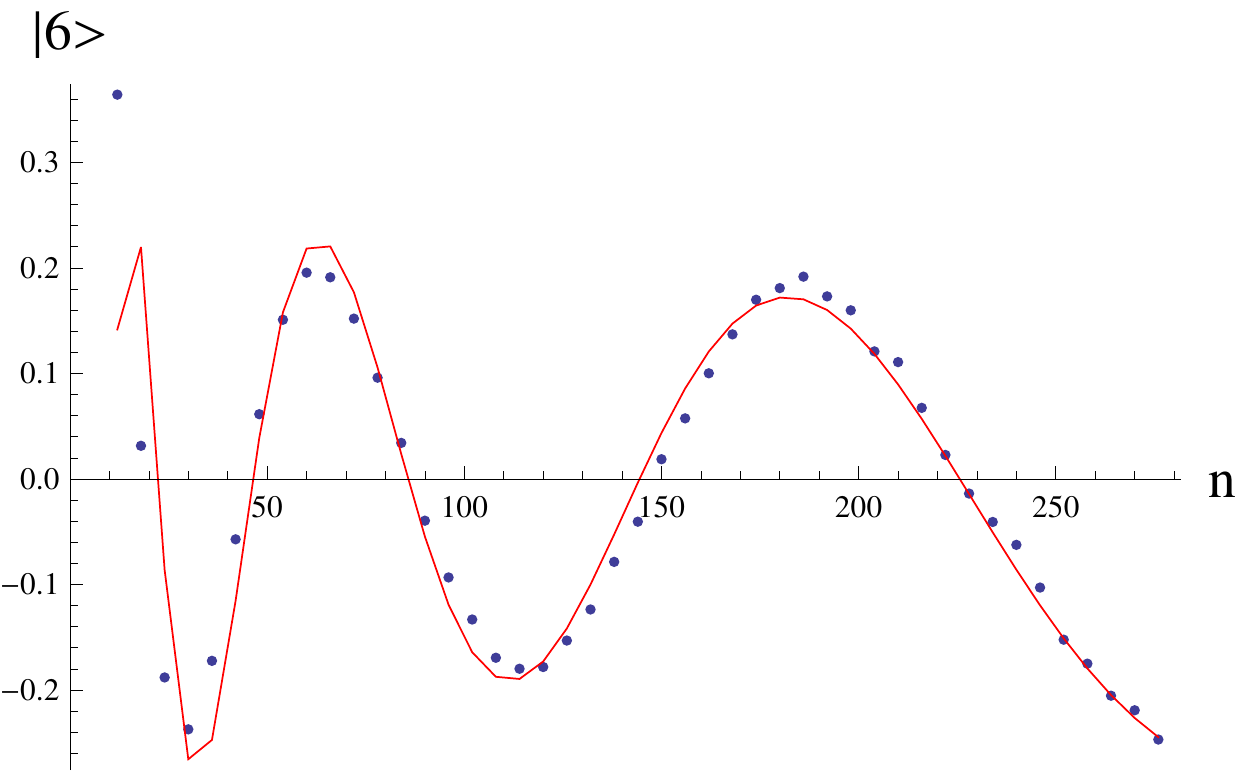}}
\caption{The first six eigenvectors of the measured $\hat M$ (blue dots) 
and ``theoretical''  matrix ${\hat M}^{(th)}$ (red line). 
${\hat M}^{(th)}$ was calculated using the effective 
action for the stalk range (\ref{Seffform}). }
\label{vec16}
\end{figure}

Having determined by a best fit the parameters of 
the effective action we can calculate the ``theoretical'' 
transfer matrix $ \hat M^{(th)}$ using (\ref{seff}) and (\ref{Seffform}).
To appreciate the quality of this approximation we present a 
plot of  six lowest eigenvalues of the measured $\hat M$ and the ``theoretical'' 
matrix ${\hat M}^{(th)}$ as well as 
the comparison of their six lowest eigenvectors. 
In each case the continuous line corresponds to the  ${\hat M}^{(th)}$ 
(Fig. \ref{eigen} -- \ref{vec16}).

\section{Discussion and conclusions.}

CDT comes with a transfer matrix $\la T | \cM|T'\ra$. 
The way CDT is defined allows us
to measure certain distributions, say $P^{t_{tot}}(n_t)$ and
$P^{t_{tot}}(n_t,m_{t+\Del t})$, of three-volumes $n_t$.
These distributions have an exact definition in
terms of the transfer matrix $\cM$ and the density matrices
$\trho(n)$ which are projections onto the subspace of 
three-dimensional combinatorial triangulations $\cT(3)$ of $S^3$
spanned by the triangulations with $n$ tetrahedra.
(see eqs. \rf{ja6}-\rf{ja8}). While the transfer matrix
is defined on the large vector space spanned by the elements
in $\cT(3)$, the actual data coming from 
Monte Carlo simulations seem to allow for a much simpler
description in terms of an ``effective'' transfer matrix $M$,
only labeled by abstract vectors $|n\ra$ referring only to
the three-volume. Not only that: basically over the whole 
range of $n_t$ the data are 
described by a transfer matrix which can be represented as 
\beql{ja40}
\la n |M| m\ra = \cN \e^{-L_{eff}(n,m)},
\eeq     
where $L_{eff}(n,m)$ is given by 
\beql{ja41}
L_{eff}(n,m) = \frac{1}{\Gamma} \left[\frac{(n-m)^2}{n \plu m \mi 2 n_0} \plu 
\mu \left(\frac{n\plu m}{2}\right)^{1/3} \mi 
\lambda \left( \frac{n\plu m}{2} \right) \plu
\delta \left(\frac{n\plu m}{2}\right)^{-\rho}\right] ,
\eeq
and with a corresponding effective action
\beql{ja42}
S_{eff} = \sum_t L_{eff} (n_t,n_{t+1}).
\eeq
The last term in \rf{ja41} unfortunately is not very well determined.
For large $n_t$ we can not really observe it. The first terms seem
to fit that data perfectly with the present statistics. For small
$n_t$ one {\it can}  detect a term like 
$\delta \left(\frac{n\plu m}{2}\right)^{-\rho}$, but as mentioned, still
$\rho$ is not well determined, and in addition the value we obtain depends 
on the specific merging of the three different distributions
one observes for small $n_t$.
Thus we cannot really claim that we have 
a result which is discretization independent. We are caught in an
unfortunate dilemma: we want to go to small $n_t$ in order
to observe this term, which indicates corrections to the 
simplest minisuperspace action. However, taking $n_t$ small 
also brings us into the region where discretization effects 
are likely to  be important. It is possible that one can find a 
window where $n_t$ is small enough for the term to be observed
via high statistic measurements, but where $n_t$ is large enough
to avoid discretization effects, but we have not yet pursued this 
in a systematic way. 

This discussion highlights an important advantage
of the present method: since $t_{tot}$ is small, we are effectively 
simulating much smaller systems than in the traditionally 
``full'' CDT computer simulations. In this way we {\it can}
actually obtain measurements of high statistics with relatively 
moderate computer resources and in a finite amount of time.

The amazing accuracy with which the effective 
transfer matrix seems to  be described by eq.\ \rf{ja40} 
indicates that one obtains a good approximation to 
the partition function by writing:
\beql{ja43}
\cZ_{t_{tot}} = \tr M^{t_{tot}} = \sum_{n_{t_i}} \e^{-S_{eff}[n_{t_i}]},
\eeq
where $S_{eff}[n_{t_i}]$ is the effective action \rf{ja41}-\rf{ja42}. We have
strictly speaking only shown that this expression is a good
approximation for some special values of the bare coupling constant
of the Einstein-Hilbert action $S_R$, given by eq.\ \rf{Sdisc}.
However, without much doubt any choice of the bare coupling 
constants which places us well inside the so-called de Sitter phase
will allow for a description in terms of an $S_{eff}[n_{t_i}]$, just with 
different $\Gamma$, $\mu$ and $\lam$. Let us assume that this 
is also true in the two other phases which have been observed 
in the ``full'' CDT theory. If this is the case one can actually
use $S_{eff}[n_{t_i}]$ and the expression \rf{ja43} to study the 
phase structure of CDT. This is of course much easier than using
the full system. This is precisely what has been done in a recent paper 
\cite{burda}. Seemingly one obtains   a good 
qualitative description of the CDT phase diagram (and 
also new interesting phase structures), corroborating the conjecture
that the functional form \rf{ja41}-\rf{ja42} might be sufficient 
to describe CDT for all choices of the bare coupling constants of the 
Einstein-Hilbert action \rf{Sdisc}. Checking this is an 
obvious task for the future. However, an even more interesting 
application of the multi-canonical Monte Carlo simulation method 
developed here is that it might allow us to investigate the CDT 
phase transitions in more detail. A possible UV scaling limit 
of the CDT theory has to be associated with these phase transitions.

\vspace{1cm}
\noindent
{\bf Acknowledgments}\\
One of the authors (JJ) thanks the Polish Ministry of Science Grants No. N N202 22913 and 182/N-QGG/2008/0.
JA would like to thank the Institute of 
Theoretical Physics and
the Department of Physics and Astronomy at Utrecht University for hospitality 
and financial support. JA also thanks the Perimeter Institute for hospitality
and financial support. JA and AG thank the Danish Research Council 
for financial support via the grant ``Quantum gravity and the role
of Black holes''. 
AG acknowledges a partial support by the Polish Ministry of 
Science grant N N202 229137 (2009-2012).

\end{document}